\documentclass[]{elsarticle}
\usepackage{natbib}
\usepackage{aas_macros} 
\usepackage{graphicx}
\usepackage{url}
\usepackage{rotating}

\journal{Icarus}

\biboptions{authoryear}

\begin{document}

\begin{frontmatter}

\title{Saturn atmospheric dynamics one year after Cassini: Long-lived features and time variations in the drift of the Hexagon}

\author[Affiliation1]{R. Hueso\corref{mycorrespondingauthor}}
\cortext[mycorrespondingauthor]{Corresponding author}
\ead{ricardo.hueso@ehu.eus}
\author[Affiliation1]{A. S\'anchez-Lavega}
\author[Affiliation1]{J. F. Rojas}
\author[Affiliation2]{A. A. Simon}
\author[Affiliation3]{T. Barry}
\author[Affiliation1]{T. del R\'io-Gaztelurrutia}
\author[Affiliation4]{A. Antu\~nano}
\author[Affiliation5]{K. M. Sayanagi}
\author[Affiliation6]{M. Delcroix}
\author[Affiliation5]{L.N. Fletcher}
\author[Affiliation7]{E. Garc\'ia-Melendo}
\author[Affiliation1]{S. P\'erez-Hoyos}
\author[Affiliation4]{J. Blalock}
\author[Affiliation8]{F. Colas}
\author[Affiliation9]{J. M. G\'omez-Forrellad}
\author[Affiliation4]{J. L. Gunnarson}
\author[Affiliation10]{D. Peach}
\author[Affiliation11]{M. H. Wong}


\address[Affiliation1]{Dpto. F\'isica Aplicada I, Escuela de Ingenier\'ia de Bilbao, UPV/EHU, Plaza Ingeniero Torres Quevedo, 1, 48013, Bilbao, Spain}
\address[Affiliation2]{NASA Goddard Space Flight Center, Solar System Exploration Division, 8800 Greenbelt Road, Greenbelt, MD 2077, USA}
\address[Affiliation3]{Broken Hill Observatory, 406 Bromide Street, Broken Hill, New South Wales 2880, Australia}
\address[Affiliation4]{Department of Physics and Astronomy, University of Leicester, University Road, Leicester LE1 7RH, UK}
\address[Affiliation5]{Department of Atmospheric and Planetary Sciences, Hampton University, Hampton, VA 23668, USA}
\address[Affiliation6]{Soci\'et\'e Astronomique de France, Commission des observations plan\'etaires, Tournefeuille, France}
\address[Affiliation7]{Serra H\'unter Fellow. Escola Superior d’Enginyeries Industrial, Aeroespacial i Audiovisual, UPC, Terrasa, Spain}
\address[Affiliation8]{IMCCE, Observatoire de Paris, PSL Research University, CNRS-UMR 8028, Sorbonne Universités, UPMC, Univ. Lille 1, F-75014, Paris, France}
\address[Affiliation9]{Fundació Observatori Esteve Duran, Barcelona, Spain}
\address[Affiliation10]{British Astronomical Association, Burlington House, London, UK}
\address[Affiliation11]{University of California at Berkeley, Astronomy Department Berkeley, CA 947200-3411, USA}
\begin{abstract}


We examine Saturn's atmospheric dynamics with observations in the visible range from ground-based telescopes and Hubble Space Telescope (HST). We present a detailed analysis of observations acquired during 2018 obtaining drift rates of major meteorological systems from the equator to the North polar hexagon. A system of polar storms that appeared in the planet in March 2018 and remained active with a complex phenomenology at least until September is analyzed elsewhere [S\'anchez-Lavega et al., A complex storm system and a planetary-scale disturbance in Saturn’s north polar atmosphere in 2018, {\em{Nat. Ast.}}, {\em{submitted}}, 2019].  Many of the regular cloud features visible in 2018 are long-lived and can be identified in Saturn images in 2017, and in some cases, for up to a decade using also Cassini ISS images. Without considering the polar storms, the most interesting long-lived cloud systems are: 

{\em{i)}} A bright white spot in the Equatorial Zone that can be tracked continuously since 2014 with minimal changes in its zonal velocity, which was $444.3\pm3.1$ ms$^{-1}$ in 2014 and $452.4\pm1.7$ ms$^{-1}$ in 2018. This velocity is remarkably different from the zonal winds at the cloud level at its latitude during the Cassini mission, and is closer to zonal winds obtained at the time of the Voyagers flybys and to zonal winds from Cassini VIMS infrared images of the lower atmosphere. 

{\em{ii)}} A large long-lived Anticyclone Vortex, here AV, that formed after the Great White Spot of 2010-2011. This vortex has changed significantly in visual contrast, drift rate and latitude with minor changes in size over the last years. 

{\em{iii)}} A system of subpolar vortices at latitudes 60-65$^{\circ}$N present at least since 2011. These vortices and additional atmospheric features here studied follow drift rates consistent with zonal winds obtained by Cassini. 

We also present a study of the positions of the vertices of Saturn's North polar hexagon from 2015 to 2018. These measurements are compared with previous analyses during the Cassini mission (2007-2014), observations with HST in the 90s, and data from the Voyagers in 1980-1981 to explore the long-term variability of the hexagon's drift rate. We find variations in the drift rate of the hexagon through these epochs that can not be fit by seasonal changes in the polar area. Instead, the different drift rates reinforce the role of the North Polar Spot that was present in the Voyager epoch and in the early 90s to cause a faster drift rate of the hexagon at that time compared with the current slower one.


\end{abstract}

\end{frontmatter}


\section{Introduction}

Saturn's atmosphere has bands, zonal jets, wave structures and vortices that resemble similar phenomena on Jupiter \citep{delGenio2009}. Both planets rotate rapidly and have internal heat sources. Saturn also has seasons, but the temporal coverage of atmospheric phenomena on Saturn from spacecraft observations does not cover the seasonal cycle to study a full Saturn year. The Cassini Mission observed the Saturn system from 2004 to 2017 from early-mid North winter to summer solstice, covering half a Saturn year and resulting in an unprecedented observational record of its atmosphere \citep{Showman2018}. The mission ended in September 2017 by plunging into Saturn's atmosphere \citep{Edgington2016, Spilker_Cospar2018}. Without Cassini, ground-based telescopes and observations from space telescopes such as Hubble Space Telescope (HST) and the future James Webb Space Telescope (JWST) \citep{Norwood2016} can be used to investigate seasonal and non-seasonal changes in Saturn's atmosphere. 

Obtaining information about atmospheric dynamics requires frequent observations over extended periods of time that are difficult to obtain from highly competitive facilities. Amateur astronomers observe Saturn very often, and the use of fast acquisition cameras, atmospheric dispersion correctors, and image processing software has resulted in significant improvements in the quality of amateur observations of solar system planets \citep{Mousis2014}. For Saturn, amateur observations have been used to study the global atmospheric dynamics, winds and storms in the planet before Cassini \citep{Sanchez-Lavega2004}, determine the long-term evolution of Saturn's Great White Spot (GWS) in 2010-2011 \citep{Sanchez-Lavega_Nature2011, Sanchez-Lavega_Icarus2012}, and study the onset and evolution of convective storms in the storm-alley discovered by Cassini \citep{Delcroix2010, Dyudina2007, Fischer2019}. Non-convective, and in many cases more subtle, features can also be observed in amateur images. For instance, \citet{Sanchez-Lavega_NatCom2016} used amateur and HST images to investigate the evolution of fast moving equatorial features over 2014 and 2015, and \citet{delRio-Gaztelurrutia2018} studied with similar observations the onset and evolution of a large polar perturbation caused by the interaction of a system of subpolar vortices. 

In late March 2018 the onset of a polar storm that emerged at planetographic latitude ($66.9 \pm 0.9$)$^{\circ}$N was first observed by M. Sparrenberger, an amateur astronomer from Brazil. The storm spread zonally over months and developed a second bright spot later around mid-May at planetographic latitude ($69.2 \pm 1.1$)$^{\circ}$N. 
These storms ignited the interest of amateur astronomers that observed Saturn regularly. 
The full phenomenology of the polar storms was very complex and is covered in detail by \citet{Sanchez-Lavega2019}. Here we focus on the study of the overall regular atmosphere of the planet and only brief mentions to the polar storms will be made.


Our goal in this paper is to extend in time our knowledge of Saturn's atmosphere beyond the Cassini mission. For that purpose we use a combination of frequent amateur observations, several observations at 1-2.2 m telescopes and a set of observations acquired with  the Hubble Space Telescope (HST) as part of its Outer Planets Atmospheres Legacy (OPAL) program.
We place particular attention in extending previous studies of the longest-lived atmospheric features that are still observable in the planet. These include the bright white spot (WS) in the Equatorial Zone (EZ) that appeared in the planet in 2014 and that has been observed since then \citep{Sanchez-Lavega_NatCom2016}, the large Dark Anticyclonic Vortex AV that formed after the onset of Saturn's Great White Spot in 2010-2011 \citep{Fletcher2011, Sanchez-Lavega_Icarus2012, Sayanagi2013}, the set of subpolar vortices formed by the Ancicylone-Cyclone-Anticyclone (ACA) system and a single vortex southwards of it \citep{delRio-Gaztelurrutia2018}. We also extend previous results about the long-term evolution of Saturn's North polar hexagon \citep{Sanchez-Lavega2014}. For some features of particular interest we also explore their long-term evolution using images obtained by the Cassini ISS instrument.

\begin{figure}[htp]
   \centering
   \includegraphics[width=12.50cm]{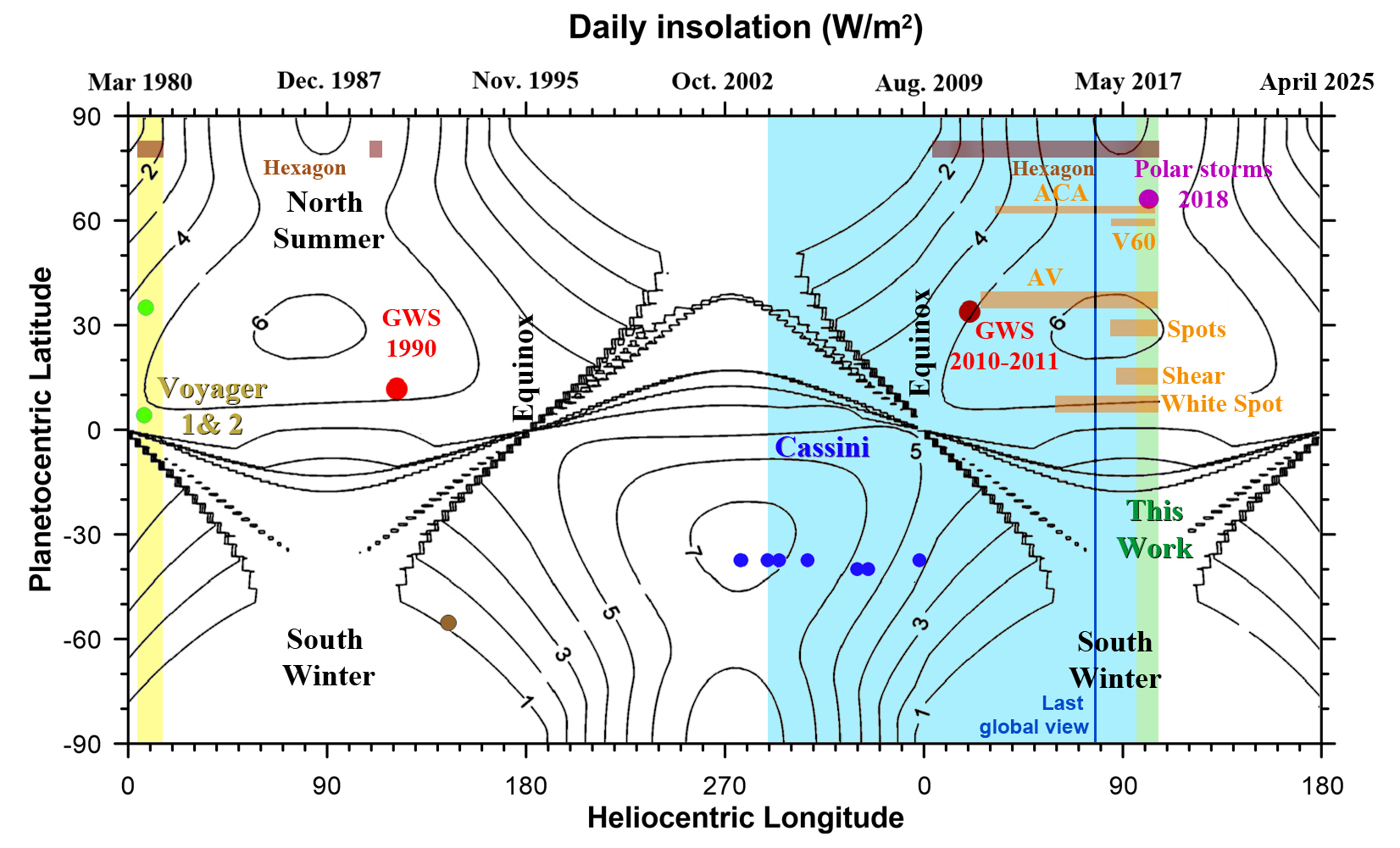}
   \caption{Daily insolation in Saturn as a function of time and latitude. The insolation has been calculated following \citet{Brinkman1979} and \citet{Perez-Hoyos2006} and incorporates sunlight reflections from the rings. Vertical colored regions show the time span covered by Voyagers 1 and 2 (yellow shaded region), Cassini (blue shaded region) and the current work (green shaded region). Dots show the location in latitude and time of convective storms in the planet (green and blue for small storms, red for giant storms or Great White Spot (GWS) and magenta for the polar storms in 2018 discussed by \citealt{Sanchez-Lavega2019}). Storms data for this figure have been adapted from \citet{Sanchez-Lavega2019}. Horizontal bars show the latitude and period of time of study for the atmospheric features discussed in this work. A vertical blue line shows the last global view of the planet obtained by Cassini.}
   \label{fig:figure_seasons}
    \end{figure}

Figure \ref{fig:figure_seasons} displays the seasonal context for the cloud systems here studied. We show the solar insolation on top of the atmosphere as a function of time, solar longitude (Ls) and latitude in Saturn, and we compare the period of time for different atmospheric features analyzed in this work with the time ranges covered by space missions. 

The structure of this paper is as follows. We present the observational data in section 2, we show the analysis of atmospheric features in section 3, including drift rates over 2018 and for some features over several years. Among these cloud systems are short-lived features (a few weeks), long-lived ones (from one year to one decade), and the vertices that define Saturn's north polar hexagon for almost 4 decades. In section 4 we compare zonal drifts of these features with the zonal winds in Saturn and examine their possible origin. In section 5 we present an analysis of the hexagon's drift rate over the years. We summarize our findings in section 6. Unless otherwise specified, all latitudes in this paper are planetographic.

\section{Observations}

\subsection{Casssini ISS observations}

\begin{sidewaysfigure}
   \centering
   \includegraphics[width=19.50cm]{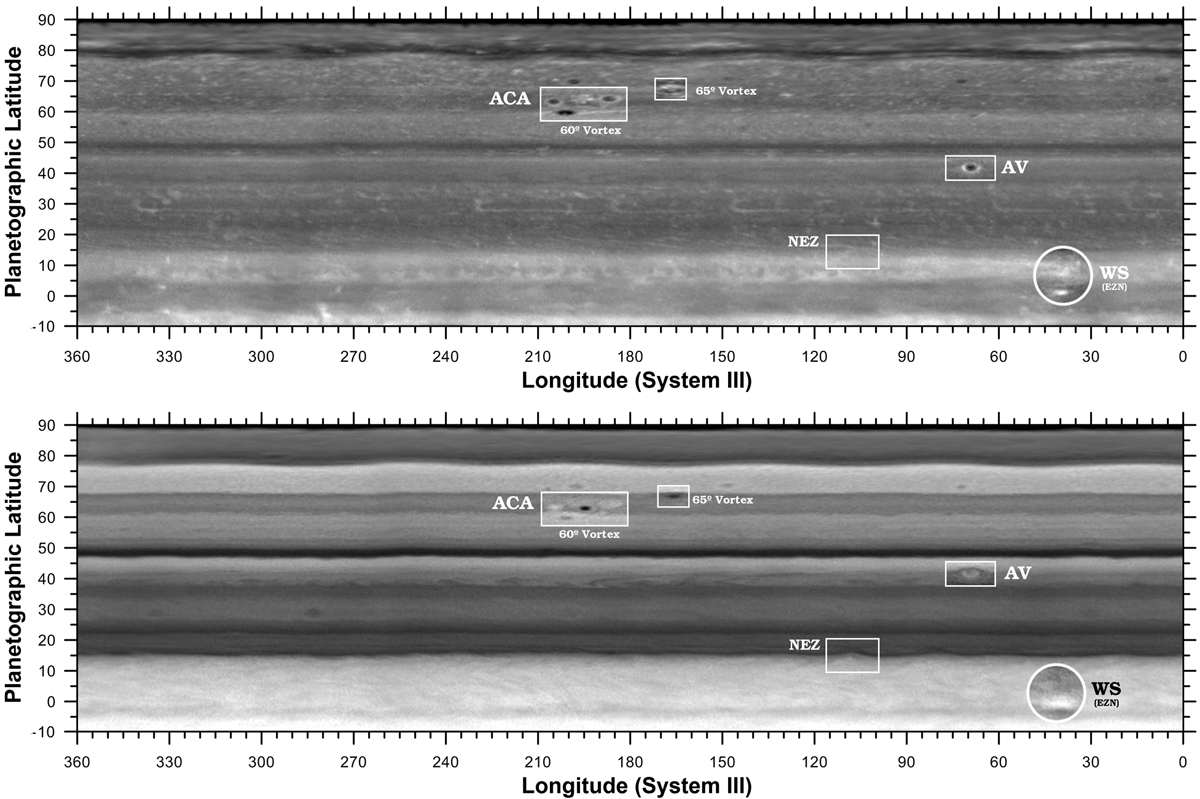}
   \caption{Saturn in 25 April 2016. Maps are based in images acquired at a spatial resolution of 350 km/pix and with a phase angle of 55$^{\circ}$. Top panel shows a map in the CB2 filter (750 nm) sensitive to the cloud level. The bottom pannel shows a map in the MT2 filter in the weak methane absorption band at 727 nm. Some of the long-lived features observable from the ground and discussed over the paper are highlighted (ACA stands for Anticyclone-Cyclone-Anticyclone; AV stands for Anticylone Vortex, WS stands for White Spot). The maps are constructed from a combination of images whose limb-darkening effects have been corrected using a Lambert law, contrast enhanced and high-pass filtered to better show the faint features. Cyclones and anticyclones are generally dark and bright respectively in images acquired in the MT2 filter.} 
   \label{fig:figure_Cassini_Map}
\end{sidewaysfigure}

The Cassini ISS instrument obtained its last global maps of the planet in April 2016. Later observations were obtained from higher latitudes as the mission prepared for its Grand Finale in September 2017 and did not achieve global coverage of the illuminated North hemisphere. Figure \ref{fig:figure_Cassini_Map} shows maps of Saturn obtained in 25 April 2016. The maps were obtained with the CB2 and MT2 filters, which are respectively sensitive to pressure levels around 350 mbar, within the tropospheric haze, and to the haze tops, closer to the tropopause \citep{Perez-Hoyos2006b, Roman2013}. The maps in this figure are shown to provide context for interpreting the atmospheric features that will be detailed later. Features of interest visible in the period 2016-2018 are highlighted. Additionally, the CB2 maps can be compared with Cassini ISS maps of the North hemisphere published by \citep{Trammell2016} from 2008 to 2015 establishing an extense time-line of the morphology of the main features visible in the planet.

\subsection{Amateur observations with small telescopes}

We used amateur observations of Saturn that are available in open databases: PVOL (Planetary Virtual Observation and Laboratory, \url{http://pvol2.ehu.eus}) and ALPO (Association of Lunar an Planetary Observers) Japan (\url{http://alpo-j.asahikawa-med.ac.jp}). PVOL is a database integrated in the Virtual European Solar and Planetary Access (VESPA) \citep{Erard2018} and allows complex searches over the data \citep{Hueso_PVOL_2018}. We have explored the ensemble of amateur observations of Saturn obtained in 2018 available on PVOL and ALPO Japan and compared these images with a selection of the best images of Saturn obtained in 2017. 

   \begin{figure}[htp]
   \centering
   \includegraphics[width=12.50cm]{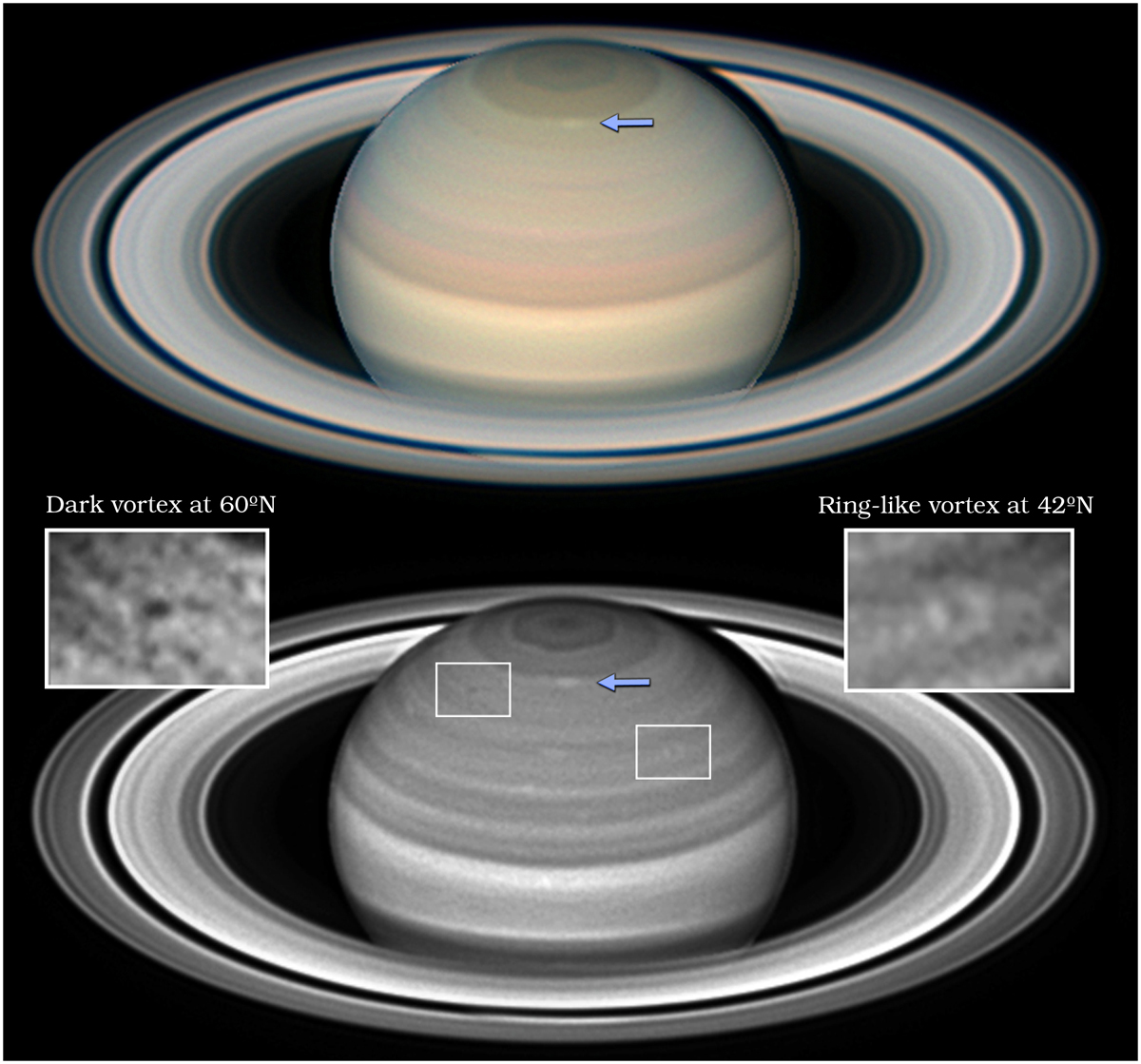}
   \caption{Example of amateur observations obtained in 2018. Images from T. Olivetti acquired in 11 April 2018. The color image on the top shows the polar storm that started in March 2018 on its first stages of development. The image on the bottom shows the results of strongly processing the red channel of the image above showing many distinguishable cloud systems that were very faint in the original. Zooms over two atmospheric vortices, the mid-latitudes AV and a subpolar vortex at 60$^{\circ}$N are highlighted. An arrow shows the North polar storm in its early stages \citep{Sanchez-Lavega2019}.}
   \label{fig:amateur}
    \end{figure}

Saturn's declination over 2017 and 2018 was $\sim-22^{\circ}$ and observers in Australia, Brazil, Chile, Phillipines, Thailand and Florida supplied the best quality data for this work. Table \ref{tab:Table_amateur} summarizes the names of the key observers whose data has been fundamental for this work. The table show dates covered by each observer from March 2017 to November 2018 and the number of nights they provided data that have been used (the number of nights observed by each one is significantly larger). The analysis of images acquired in 2017 is not exhaustive, as images from 2017 were only used to confirm the long-lived characteristics of a limited number of features repeteadly observed over 2018. Data from many other observers were also used, although with a lower number of observations, or with observations of lower quality.
Most observations in this table by D. Peach correspond to observations obtained with a 1.0m remote telescope in Chile (Chilescope) using amateur cameras and processing techniques.

\begin{sidewaystable} %
\centering
\begin{tabular}{llllllll}\hline
Observer             & Location        & First$_{2017}$ & Last$_{2017}$ & N$_{2017}$ & First$_{2018}$ & Last$_{2018}$ & N$_{2018}$ \\
\hline
A. Casely             & Australia        &                     &                       &                      & 2018-04-16T        & 2018-06-22T      & 13      \\
A. Wesley            & Australia        &                     &                       &                      & 2018-04-02T        & 2018-07-22T      & 9       \\
T. Barry               & Australia        & 2017-03-01T  & 2017-11-07T    & 12                  & 2018-02-16T        & 2018-11-08T      & 32       \\
B. MacDonald      & Florida          & 2017-07-05T   & 2017-08-22T    & 6                    & 2018-05-23T        & 2018-10-21T      & 35      \\
C. Go                  & Australia       & 2017-03-18T   &                       & 1                    & 2018-04-10T        & 2018-04-29T      & 14      \\
C. Foster              & South Africa & 2017-09-28T   & 2017-11-02T    & 3                    & 2018-03-01T        & 2018-11-08T       & 30      \\
D. Peach*            & Chile            & 2017-02-25T  & 2017-04-22T    & 2                    & 2018-03-31T        & 2018-08-27T       & 19      \\ 
D. Milika               & Australia      &                     &                      &                       & 2018-04-02T        & 2018-06-30T       & 8       \\
J. Camarena         & Spain          &                     &                      &                       & 2018-06-20T        & 2018-07-18T       & 9       \\
M. Sparrenberger & Brazil           &                     &                      &                       & 2018-03-29T        & 2018-07-21T       & 3       \\
P. Miles                & Australia       &                    &                      &                       & 2018-04-01T        & 2018-04-05T       & 3       \\
E. Morales           & Puerto Rico    &                    &                      &                       & 2018-04-01T        & 2018-05-07T       & 6       \\
T. Olivetti            & Australia         & 2017-04-21T & 2017-08-13T   & 5                     & 2018-04-11T        & 2018-11-03T      & 4       \\
T. Horiuchi          & Japan            & 2017-08-24T & 2017-08-25T   & 2                     & 2018-05-17T        & 2018-06-02T      & 3       \\
T. Kumamori       & Japan            &                   &                       &                        & 2018-03-23T       & 2018-07-28T      & 9       \\
W. Martins          & Brazil             &                   &                       &                        & 2018-05-01T       & 2018-07-16T      & 6   \\    
\hline
\end{tabular}
\caption{Main amateur observers, periods covered by each one and number of observations used in this work. (*) Many of the observations from this observer, but not all, were obtained with the 1.0m Chilescope.}
\label{tab:Table_amateur}
\end{sidewaystable}

Observations  in Table \ref{tab:Table_amateur} can be retrieved from the PVOL database by searching through the ``observer" field in the web search form. Figure \ref{fig:amateur} shows examples of some of the best observations. These images are obtained through a process of fast acquisition and stacking of the best frames condensing video observations obtained over a minute or two into a single image. Images or videos obtained over longer time spans are combined through a derotation process in which each image or video is navigated and the rotation of individual images is compensated through the use of appropiate software (amateur observers use the WinJupos software and derotate data acquired from a few to 50 minutes). The result is a high signal-to-noise ratio image that amateurs process using a variety of sharpening techniques that includes wavelets, high-pass filters and color compositions. The images available in PVOL and ALPO Japan are the outcome of such image processing techniques. However, for the  work presented here, this level of sharpening is generally not enough, and most of the observations have been further sharpened using a combination of high-pass filters suitable for each individual image. 

\subsection{1.05-2.2-m telescopes}

   \begin{figure}[htp]
   \centering
   \includegraphics[width=12.50cm]{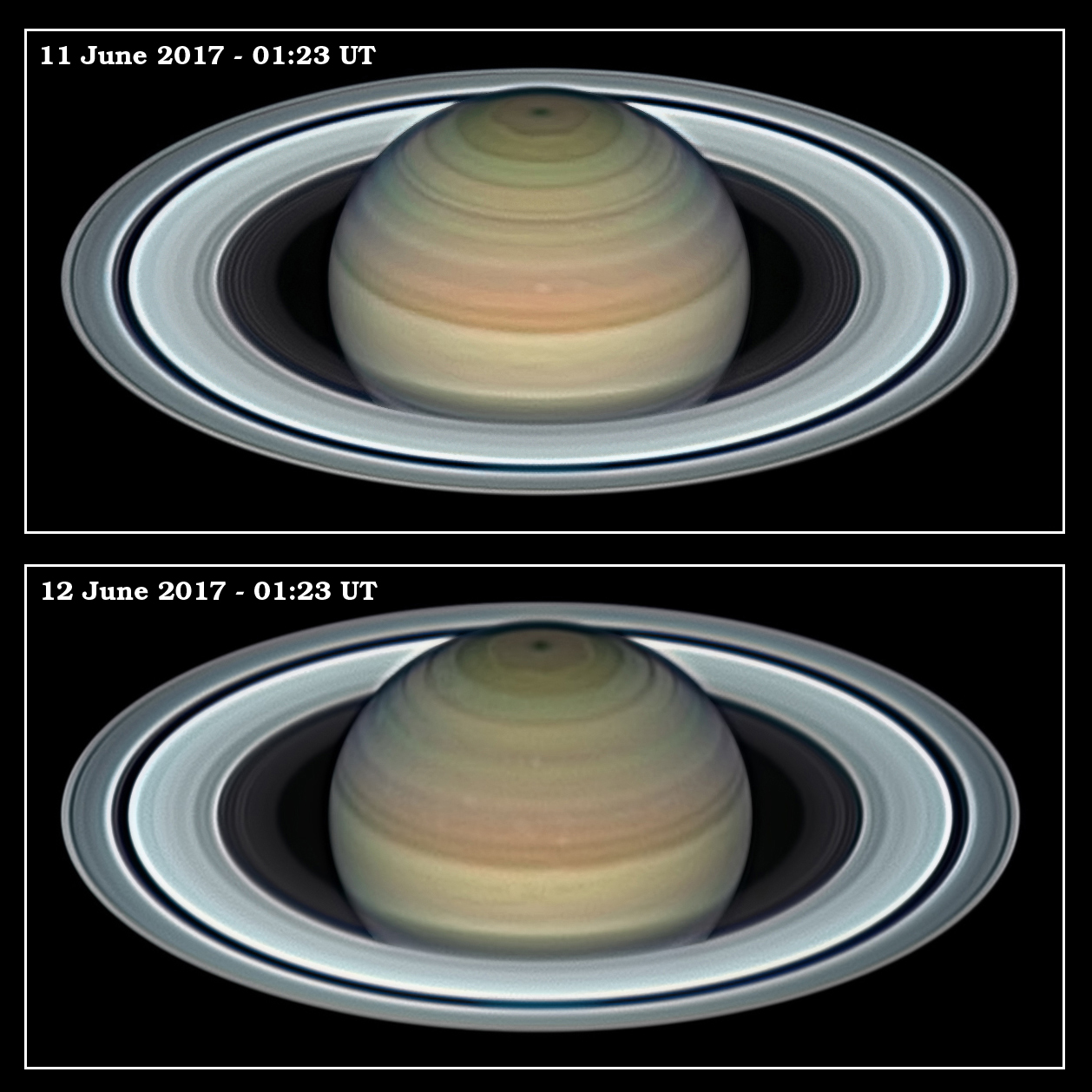}
   \caption{Saturn observations obtained with the 1.05-m telescope at Pic du Midi. Videos obtained over tens of minutes were combined using derotation techniques to build the final images. The bottom image used information from videos obtained over 46 minutes. Saturn was at elevations lower than 30$^{\circ}$ during these observations.}
   \label{fig:figure_Pic}
    \end{figure}

   \begin{figure}[htp]
   \centering
   \includegraphics[width=12.50cm]{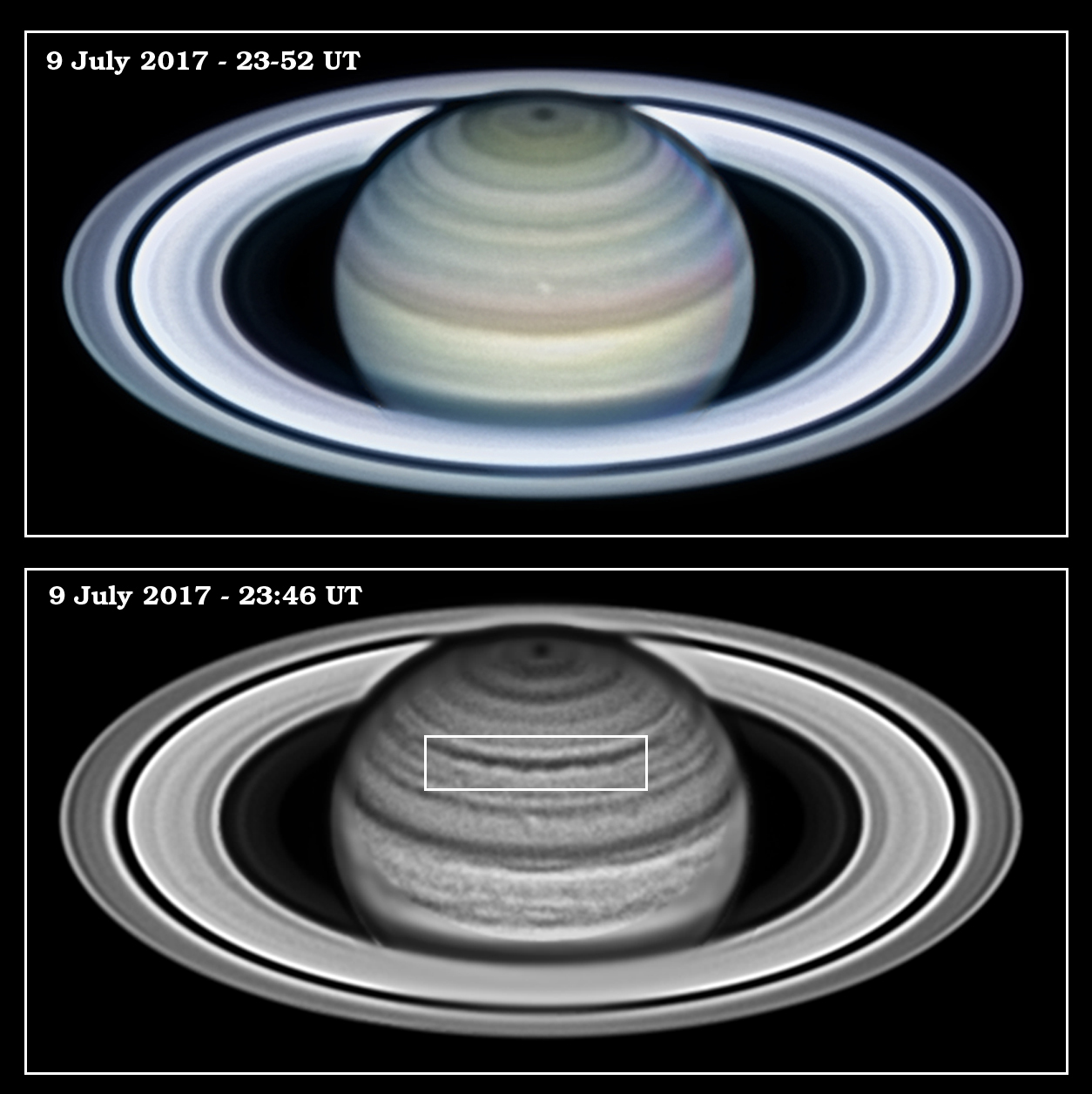}
   \caption{Examples of Saturn observations obtained with the 2.2m telescope at Calar Alto on 9 July 2017 with the PlanetCam instrument. The image in the top is a colour-composite while the image in the bottom was obtained in the SWIR chanel in the 1.0-1.7 $\mu$m wavelength (RG1000 filter in the PlanetCam instrument). In both cases, consecutive images have been combined into a single final image using derotation techniques. The top panel is the result of derotation over images acquired over a period of 15 minutes.  The bottom panel is the result of extreme derotation over images in the RG1000 filter acquired over a period of 1.5hrs obtaining 20 videos of 1 minute resulting in 60,000 frames from which the best 6,000 frames are combined and derotated. The wavy structure in this image at 47$^{\circ}$N highlighted with a box is the ribbon-wave (see section \ref{section:ribbon}). Both images in this figure were acquired simultaneously with the two channels of the PlanetCam instrument at a planet elevation of 30$^{\circ}$.}
   \label{fig:figure_CAHA}
    \end{figure}

The 1.05m planetary telescope at the Pic du Midi observatory in France regularly obtains high-resolution observations of the planets using fast acquisition cameras in the visible domain. Saturn observations in 2017 were acquired on 10-12 June by the ``PicNet'' team using amateur cameras and image acquisition and processing techniques under exceptional seeing conditions. Figure \ref{fig:figure_Pic} shows an example of these observations. These Pic du Midi observations can be downloaded from the PVOL database at \url{http://pvol2.ehu.es}. Different observing runs were attempted in 2018 but bad weather did not allow to obtain images with this quality.

We also obtained Saturn observations from the 2.2m telescope at Calar Alto using the PlanetCam UPV/EHU instrument \citep{Mendikoa2016}. This instrument allows to observe simultaneously in two wavelength ranges (visible from 400 nm to nearly 1 $\mu$m and Short-Wave InfraRed (SWIR) from 1.0 $\mu$m to 1.7 $\mu$m). Several high-quality observations were obtained in 2017 and are reported by \citet{Mendikoa2017}. However, our observations in 2018 were affected by bad weather and only some observing runs produced images that could be used in this work. Representative observations from 2017 are shown in figure \ref{fig:figure_CAHA}. 
Table \ref{tab:Pic_CAHA} shows the dates of Saturn observations obtained with these telescopes that have been used here.

\begin{table} %
\centering
\begin{tabular}{l l}\hline
Telescope         &   Dates                       \\
\hline
1.05m Pic         &   10-12 June 2017        \\
\hline
2.2m Calar Alto  &   14     June 2017\\
                       &   07     July 2017\\
                       &   23     May 2018\\
                       &   27     June 2018\\
                       &   04-05 September 2018\\ 
\hline 
\end{tabular}
\caption{Dates of succesful observations. Observations were obtained also in other dates not listed here but under worse atmospheric conditions resulting in no visible features on Saturn's disk comparable to those in these images.}
\label{tab:Pic_CAHA}
\end{table}

\begin{sidewaysfigure}
\centering
   \includegraphics[width=19.50cm]{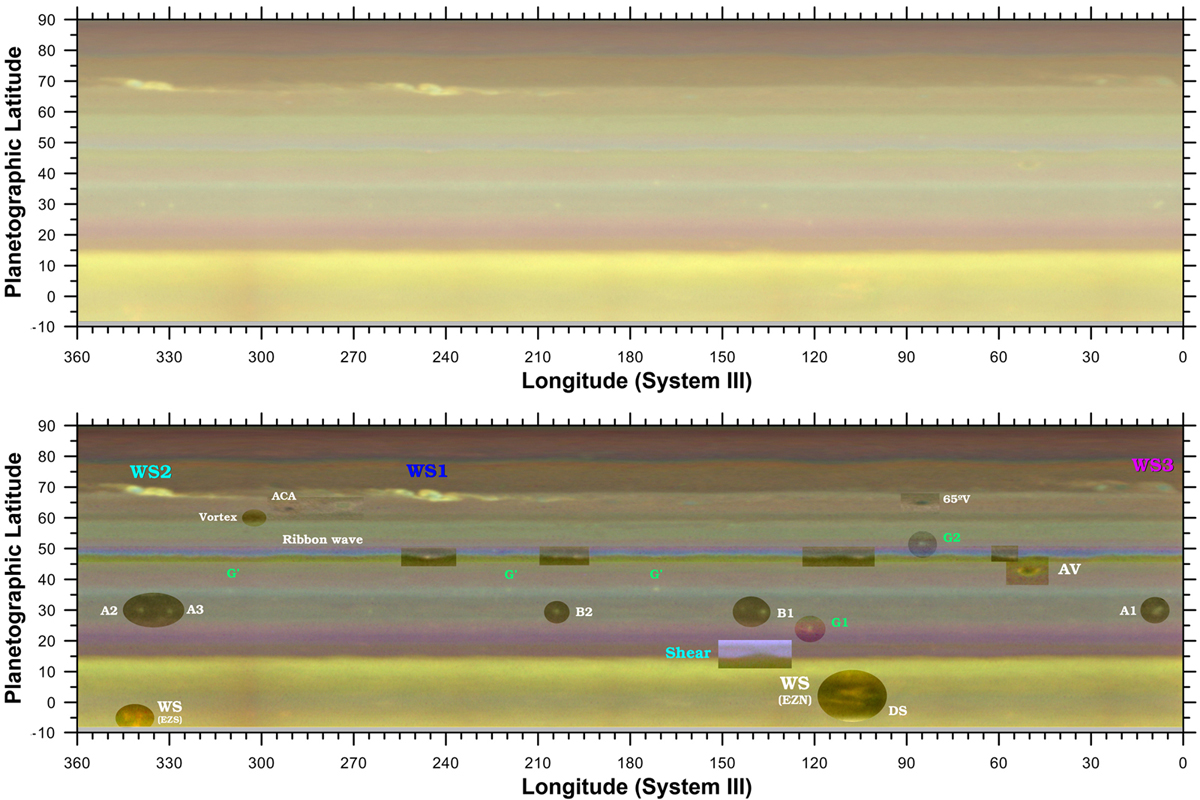}
   \caption{Saturn map from HST images acquired on 6 June 2018. The upper panel shows a color map almost unprocessed. The bottom panel shows a highly processed version of the map and highlights cloud features. Convective storms in the polar region were fully developed at the time of these observations and are labelled as WS1, WS2 and WS3 in consistency with \citep{Sanchez-Lavega2019}. Most features labelled had counterparts in amateur images. Features G' were bright in green filters but did not have counterparts in the amateur record. The ``shear" feature and the ribbon wave (see section \ref{section:ribbon}) were more contrasted in images in blue wavelengths and in the FQ727N filter. Some vortices were more contrasted in the FQ727N filter.}
\label{fig:figure_HST}
\end{sidewaysfigure}

\subsection{Hubble Space Telescope observations}

HST observed Saturn as part of the OPAL program on 6 June 2018. The WFC3/UVIS instrument obtained 89 images that were used to produce two full maps of the planet in filters F225W, F275W, F343N, F395N, F467M, F502N, F631N, FQ727N and F763M covering two full planet rotations. The maps are available on the HST/OPAL webpage  (\url{https://archive.stsci.edu/prepds/opal/}) including color maps based on the F631N (red), F502N (green), and F395N (blue) maps. 

Figure \ref{fig:figure_HST} shows one of the two color maps of the planet. Most of the features highlighted are identifiable in amateur observations of the planet over extended periods of time. Some of them appear as very low contrast features in the color HST map but have a higher contrast at longer or shorter wavelengths and are accesible to amateur equipment. Some others are evident in the HST green filtered images but less contrasted in red wavelengths and resulted in only a few detections in the amateur record (features G1 and G2 in the HST map) or none (features G' in the HST map).

Previous observations of Saturn with the WFC3/UVIS instrument were obtained on 29-30 June 2015 \citep{Sanchez-Lavega_NatCom2016} and were incorporated in the analysis of some of the long-lived features.


\section{Analysis}
\begin{figure}
\centering
   \includegraphics[width=12.00cm]{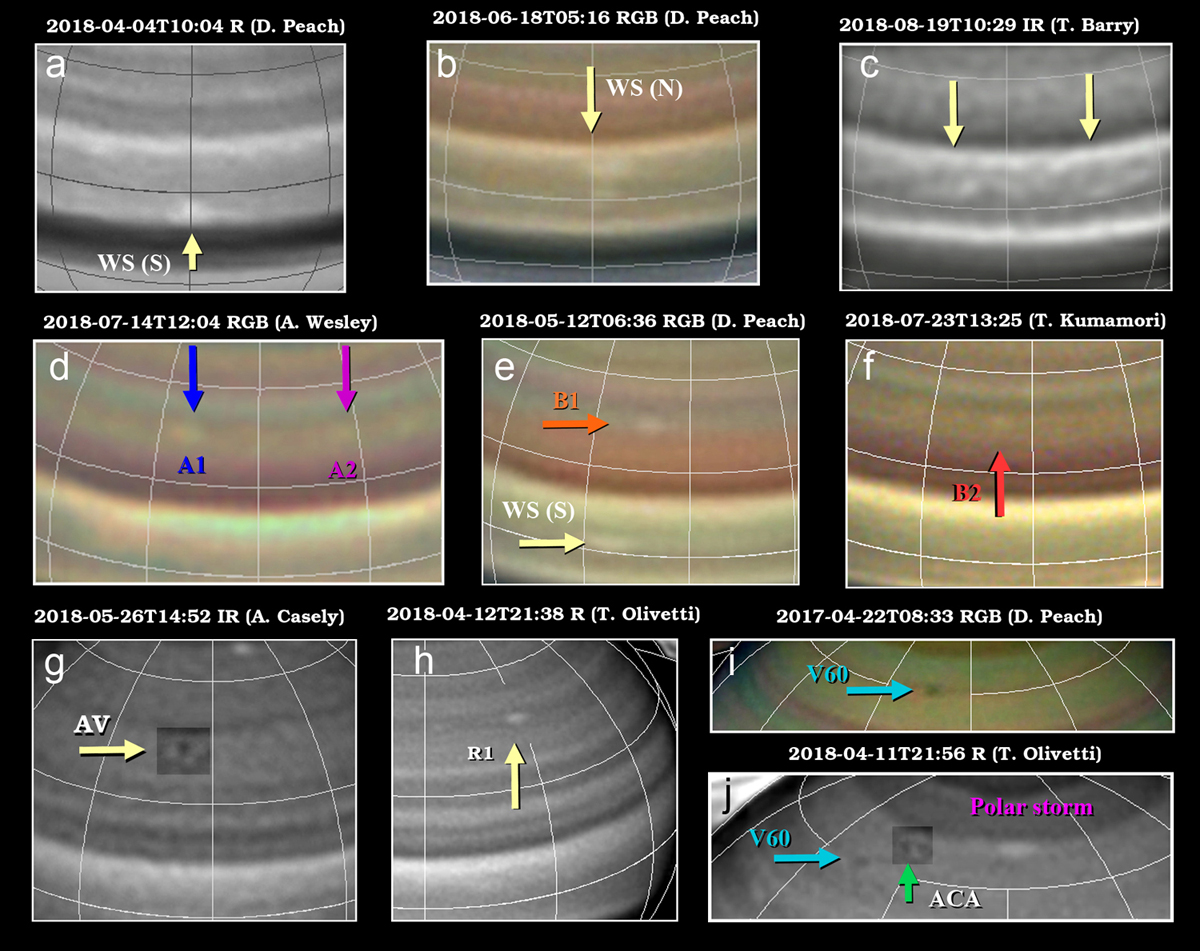}
   \caption{Selection of amateur images showing some of the atmospheric features discussed in this work. Top row and from left to right: a: South bright White Spot (WS) in the Equatorial Zone (EZ); b: bright White Spot in the North EZ; c:  elongated detached features in the EZ. Middle row: d-f: images of tropical features A1, A2, B1 and B2 at $30^{\circ}$N. Bottom panel from left to right: g: the AV vortex; h: bright features at the ribbon latittude (R1); i: vortex at $60^{\circ}$N in 2017. Panel j  shows the vortex in 2018 with the Anticyclone, Cyclone, Anticyclone (ACA) system and the nearby polar storm. Grid lines are displayed every $30^{\circ}$ or $20^{\circ}$. The features AV and ACA are extremely faint features whose contrast has been enhanced in this figure.}
   \label{fig:figure_amateur_morphologies}
\end{figure}

In this section we describe the analysis of relevant features proceeding from equatorial latitudes to the North pole. Figure \ref{fig:figure_amateur_morphologies} shows examples on amateur images of many of these features and will be referenced through the text. 

\subsection{Equatorial White Spot: 2014-2018 (Ls=58-106)}

A bright White Spot (WS) in the North Equatorial Zone (NEZ) at $6.7\pm1.6^{\circ}$N appeared in 2014  and was visible in many amateur observations acquired in 2014 and 2015. \citet{Sanchez-Lavega_NatCom2016} reported on the properties of this feature based on Cassini ISS images, ground-based data and HST images acquired on 29-30 June 2015. The bright cloud moved at 450 ms$^{-1}$. This velocity is significantly faster than zonal winds obtained from cloud tracking on Cassini ISS images on the continuum band filters through 2004-2009 ($\sim380$ ms$^{-1}$, see \citealt{Garcia-Melendo2010}). Instead, the bright WS moved at similar velocities to those of equatorial clouds observed in visible light in 1980-1981 by Voyager 1 and 2 \citep{Sanchez-Lavega2000}. The bright WS was visible again in 2016. However, it had an apparent lower contrast and it was only visible in some of the best ground-based observations of 2016. The Cassini map in Figure \ref{fig:figure_Cassini_Map} shows this feature with lower contrast than HST images acquired in 2015. Later ground-based observations in 2017 and 2018 showed a bright equatorial cloud with high contrast. In the best observations the WS had an ``S" shape with two parallel strikes of $8.2\pm1.9^{\circ}$ in longitude with a latitudinal size of $7.4\pm2.2^{\circ}$, very similar to the morphology observed in 2015 in HST images (see figure 4 in \citealt{Sanchez-Lavega_NatCom2016}). The morphology of this equatorial feature in 2017 and 2018 is similar in most of the observations with enough quality to inspect its shape (figure \ref{fig:figure_amateur_morphologies}).

   \begin{figure}[htp]
   \centering
   \includegraphics[width=12.50cm]{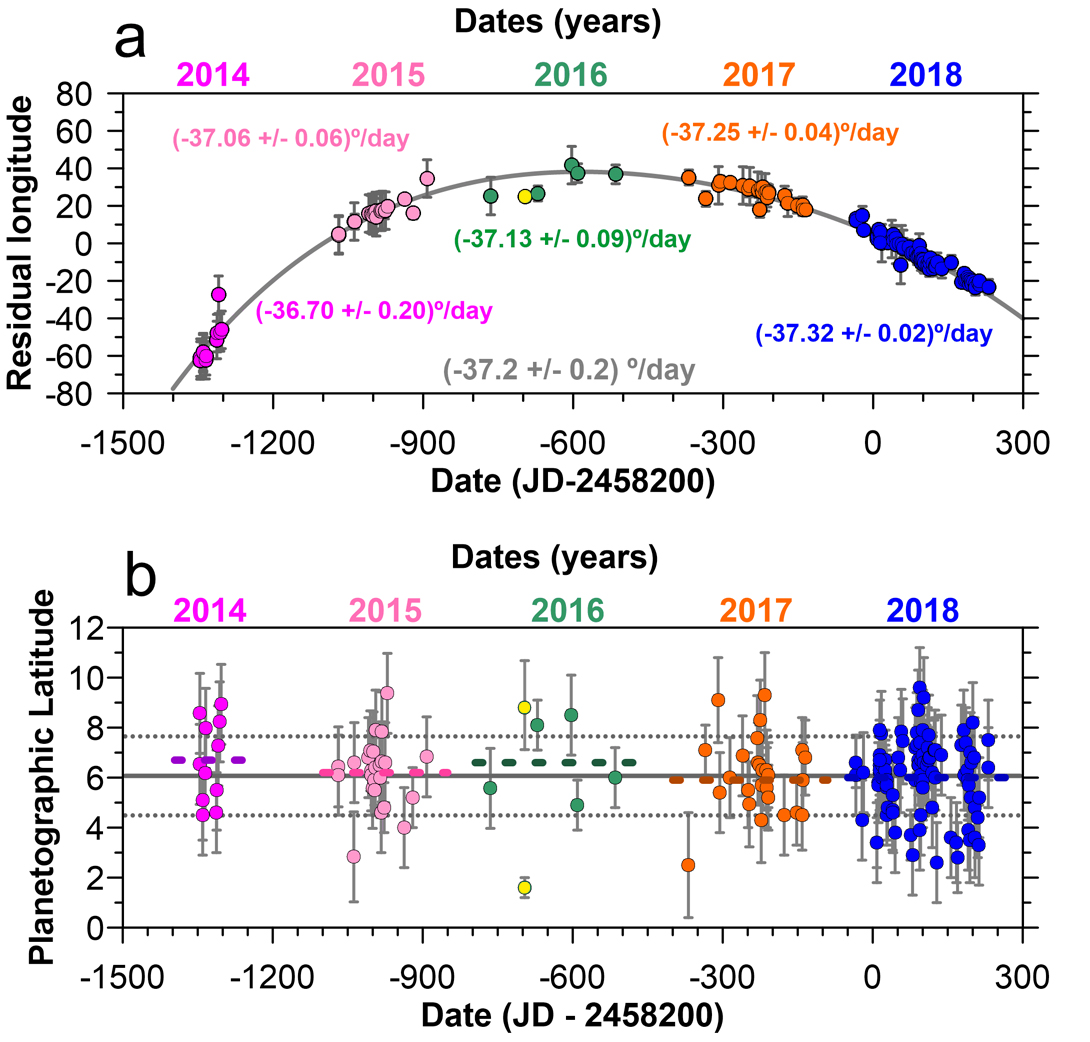}
   \caption{Long-term tracking of the White Spot in the Equatorial Zone. (a) Residual longitude resulting from a fit of all data from 2014 to 2018 with a linear drift rate of -37.19 ($^{\circ}$ day$^{-1}$). Drift rates and their uncertainty for each year are given in the figure. (b) Planetographic latitude of the WS. Their average value is shown with a continuous grey line and the uncertainty from the standard distribution of the measurements is shown with grey dotted lines. Color dashed lines correspond to the mean values of the latitude for individual years. Yellow circles in 2016 correspond to the single measurement of the White Spot in Cassini images. The two latitudes in the bottom pannel correspond to the main feature and the bright narrow feature detached from it, but moving at the same speed on the Cassini ISS images. Here and in other figures involving tracking of atmospheric features dates are given with respect to Julian Day 2458200, or 22 March 2018.} 
   \label{fig:EZ_Tracking_2018}
    \end{figure}

We tracked the position of this bright equatorial feature in system III. Then we modified the longitudes taking into account the number of full rotations around the planet performed by the WS (roughly one rotation every 9.7 days) and decreasing by 360 degrees the modified longitude for each planetary rotation in order to keep the time variation of the spot's longitude monotonic \citep{Sanchez-Lavega_NatCom2016}. These modified longitudes allow to study linear trends and drifts of the cloud systems over several years. Residual longitudes obtained by substracting a linear fit to the measurements in the modified longitude system show the goodness of the fit and can be used to expose small variations of the drift rates for individual years. Figure \ref{fig:EZ_Tracking_2018} shows residual longitudes for the analysis of the WS in the period 2014-2018 with data from 2014 and 2015 coming from \citet{Sanchez-Lavega_NatCom2016}. This long-term tracking covers 162 rotations of the bright feature over the planet's system III longitude. We consider that an uncertainty of a full rotation over the planet exists between the 2014 and 2015 data. However, the effect of this uncertainty has a very limited impact in the global drift rate derived from the data (an error in longitude equivalent to a full rotation in the period 2014-2018 would result in an uncertainty of 0.2  $^{\circ}$day$^{-1}$ for the global fit in this period). The positions shown in figure \ref{fig:EZ_Tracking_2018} are those that minimize changes in the drift rate. For individual years the data sampling is high enough to avoid having missed a single rotation around the planet, and uncertainties in the drift rates are much smaller. 

Figure \ref{fig:EZ_Tracking_2018} shows that the drift rate of this White Spot changed gradually from $444\pm 3$ ms$^{-1}$ ($-36.7\pm0.2$ $^{\circ}$day$^{-1}$) in 2014 to 
$452.4\pm 1.7$ ms$^{-1}$ ($-37.32\pm0.02$ $^{\circ}$day$^{-1}$) in 2018. Measurement uncertainties in the latitude of this feature are on the order of 1.0$^{\circ}$ for the best images in 2018 and about 1.6$^{\circ}$ for the statistical analysis of all data from 2014 to 2018. The uncertainties in latitude dominate the errors in the zonal drift in ms$^{-1}$ of the bright feature and do not allow to determine if changes in the drift rate are produced by changes in latitude. 

   \begin{figure}[htp]
   \centering
   \includegraphics[width=7.50cm]{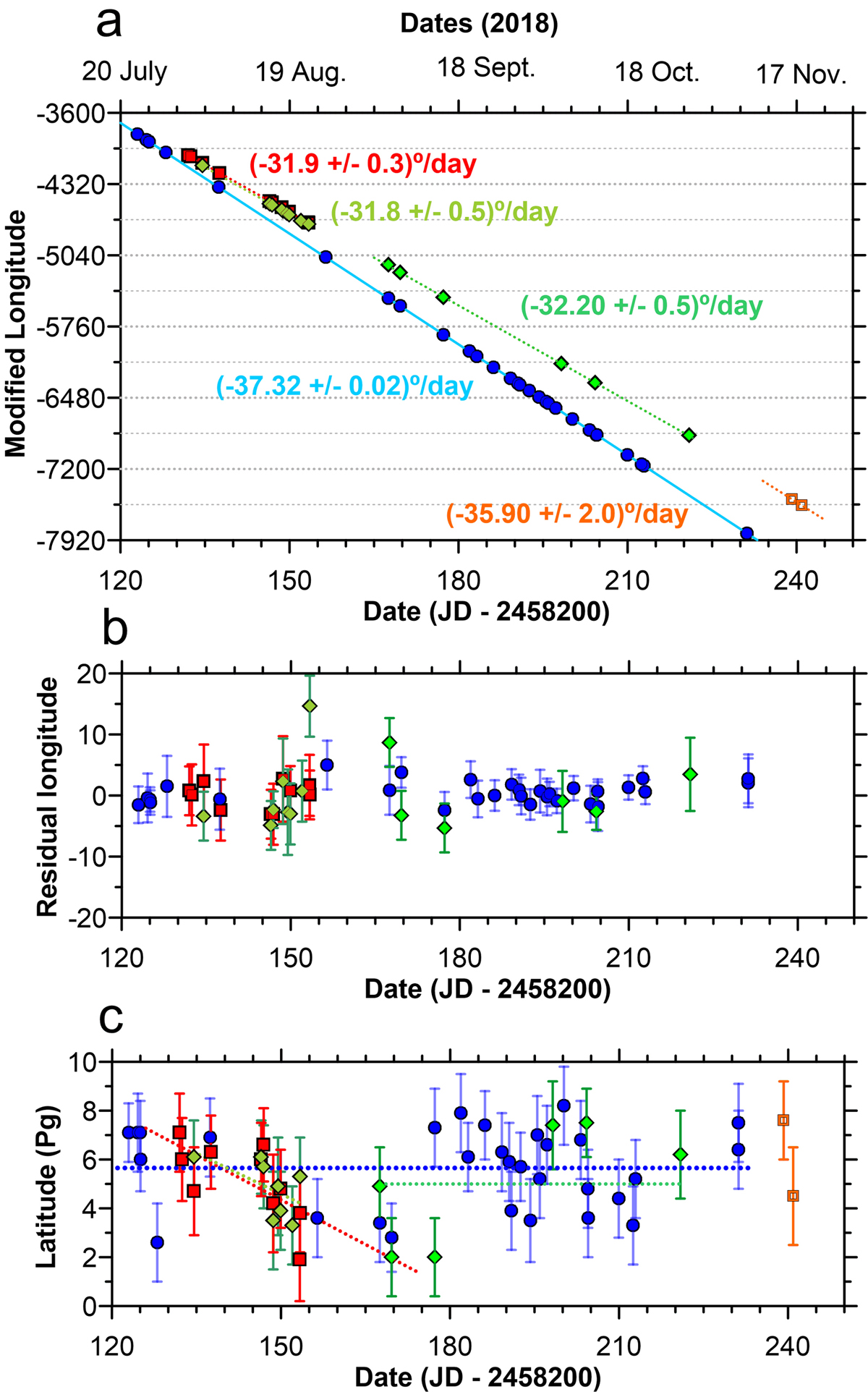}
   \caption{Tracking of equatorial features over the second half of 2018. Panel a shows longitudinal measurements of different features. The vertical axis is a modified system III longitude displaying data corresponding to different rotations of these cloud systems over the planet. Grid-lines are drawn every 360 degrees and drift rates of individual features with the uncertainties from the measurements are indicated with linear fits. Error bars in this scale are negligible. Panel b shows residual longitudes when substracting the linear drifts to the positions of each feature with error bars showing estimated errors in individual measurements. Panel c shows the latitudinal position of these features. Lines are linear fits to the latitudes. The last data points in November are only temptative as they appear in images with worse quality. In all panels, blue circles respresent the main bright white spot. Red squares and green diamonds show elongated features originating close to the main bright white spot around 24 July. Orange circles show additional equatorial clouds.}
   \label{fig:EZ_Tracking_2018B}
    \end{figure}

Over the second half of 2018 other equatorial features were visible at the same latitude. Besides the bright spot described above, in the best images there were some elongated features that drifted at different speeds. Figure \ref{fig:figure_amateur_morphologies}c shows an example of these elongated equatorial clouds. Figure \ref{fig:EZ_Tracking_2018B} shows the longitudinal positions of these features, their drift rates and their residual longitudes after substracting a linear fit to the longitudes of the individual cloud systems. As in the previous case, a modified longitude system is used to avoid sharp changes in longitude when the features travel a full planetary rotation in the system III longitudes. Two of these clouds formed very close to each other around 24 July 2018 and close to the main bright white spot described above. The two new clouds drifted more slowly in longitude separating from the main WS but remaining close to each other. Other bright clouds were visible in late August and through September and October but a definitive identification of them is not possible. 

\subsection{South Equatorial Zone: 2018 (Ls=98-106)}
   \begin{figure}[htp]
   \centering
   \includegraphics[width=10.50cm]{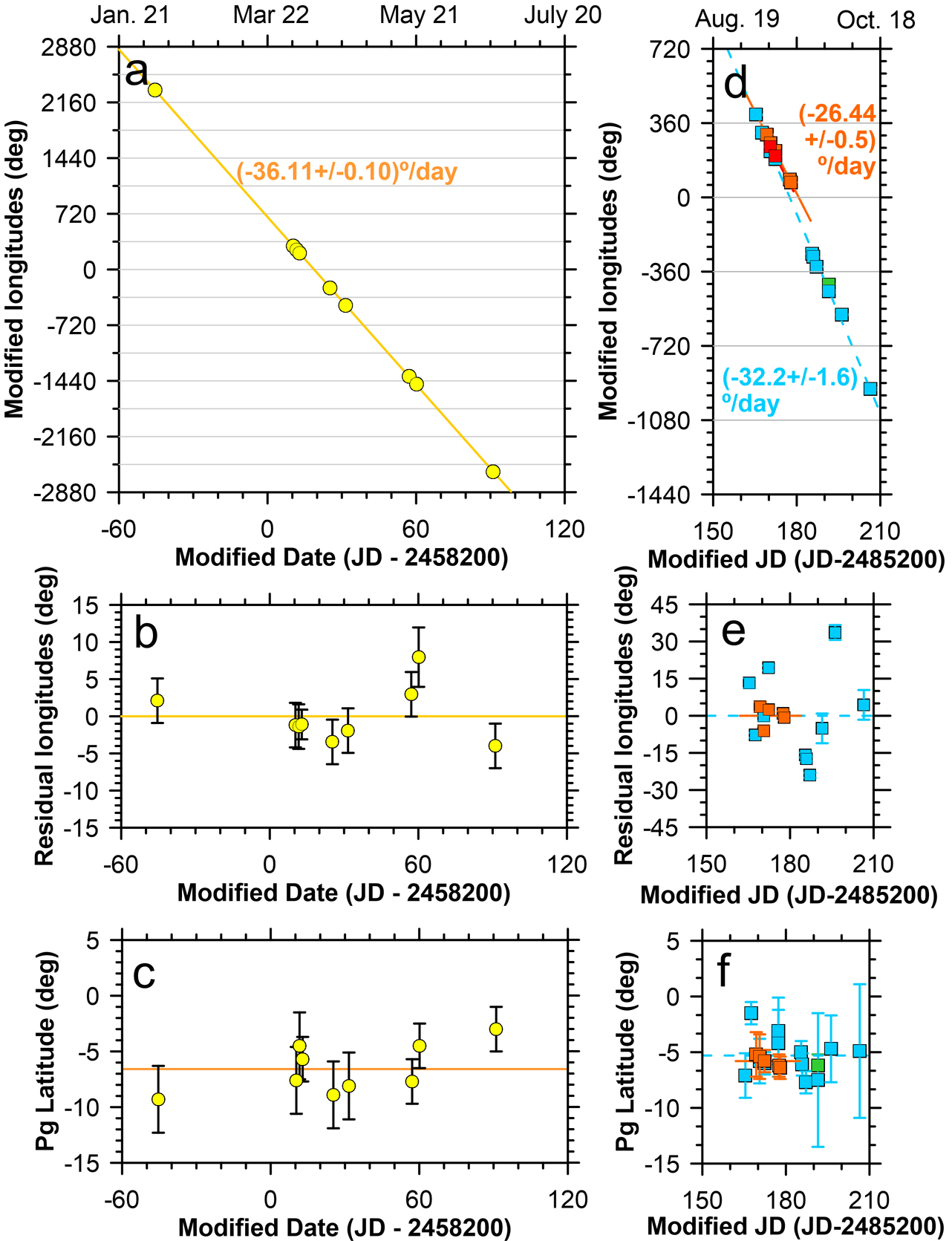}
   \caption{Tracking of South equatorial features in 2018. Panel a shows the tracking of the most conspicuous South Equatorial feature. Grid-lines are plot every 360$^{\circ}$. Panel b shows residual longitudes after substracting the linear fit to the data. Panel c shows the latitude of this feature. Panel d shows the longitudinal positions of different features observed in September-October 2018 together with possible linear fits to the two main features. Grid-lines are plot every 360$^{\circ}$. Panel e shows the residual longitudes of these details after substracting the linear drift. Panel f shows the latitudes of these features. Lines in each plot are linear fits to the data. Measurement errors in panels a, d and in most points in panel e are smaller than the symbols.}
   \label{fig:SEZ_Tracking_2018B}
    \end{figure}

In 2018 a bright cloud feature in the South Equatorial Zone (SEZ) at approximate latitude $(6.7 \pm 2.2)^{\circ}$S was also observed regularly (examples are given in Figures \ref{fig:figure_amateur_morphologies}a and Figure \ref{fig:figure_HST}). Similar features are observable in the Cassini ISS images in 2016 in Figure \ref{fig:figure_Cassini_Map}, but there are not enough data to relate the 2016 southern bright features with those observed in 2018. The detections in 2018 occured from February to June and Figure \ref{fig:SEZ_Tracking_2018B} shows the tracking of this feature resulting in a drift speed of $437.0\pm2.8$ ms$^{-1}$ ($-36.11\pm0.08$ $^{\circ}$day$^{-1}$). This bright cloud is not visible in later images in July or August. 

New bright clouds appeared at the same latitude in the SEZ in September and their tracking is shown in the right panels of figure \ref{fig:SEZ_Tracking_2018B}. A bright cloud that lasted for about 45 days held a drift rate of $391\pm6.3$ ms$^{-1}$ ($-32.2\pm2.5$ $^{\circ}$day$^{-1}$). Other features were visible over a few days after appearing close to this bright white spot. The drift rate of the longest-lived (8 days) 
was $321\pm6$ ms$^{-1}$ ($-26.4\pm0.5$ $^{\circ}$day$^{-1}$), which is significantly smaller than the drift speed of any of the previous cloud systems. These bright SEZ clouds appeared roughly at the same time as the elongated features in the NEZ. Both of them seemed smaller and more compact than the previous bright white spot in the SEZ.

\subsection{Northern edge of the Equatorial Zone: 2016-2018 (Ls=78-106)}
   \begin{figure}[htp]
   \centering
   \includegraphics[width=12.50cm]{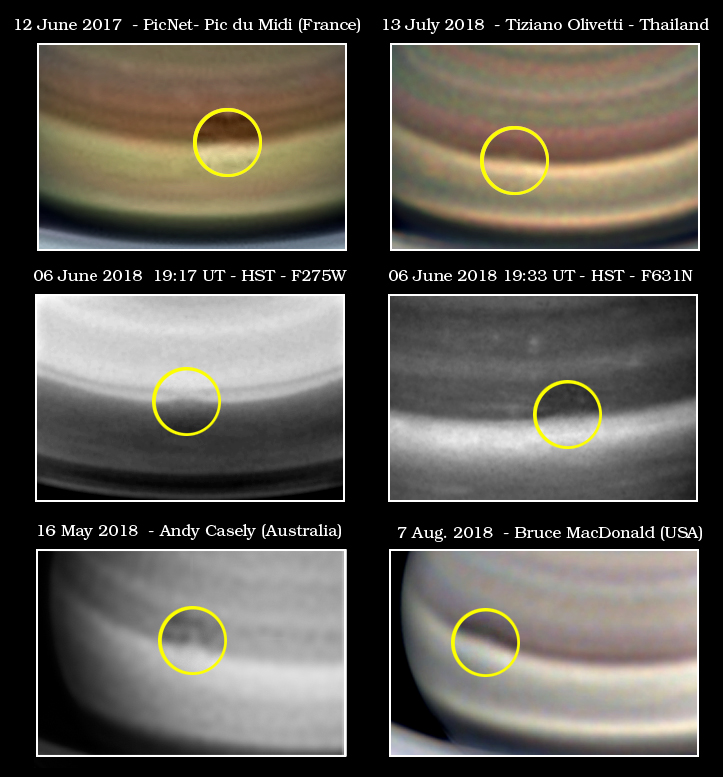}
   \caption{Observations of the shear-like feature at 15$^{\circ}$N. Top row: Images in 2017 show this faint feature in color and red wavelengths. Middle row: HST images in the center show the different contrast of this structure, which is the strongest at 275 nm and is reduced in red wavelenths at 631 nm. Bottom row: Examples of amateur observations of this feature in 2018. Yellow circles highlight the position of the feature. Due to its subtle nature, we recommend looking at the full resolution image in the electronic version of this paper.}
   \label{fig:Shear_Examples}
    \end{figure}

\begin{figure}[htp]
   \centering
   \includegraphics[width=9.00cm]{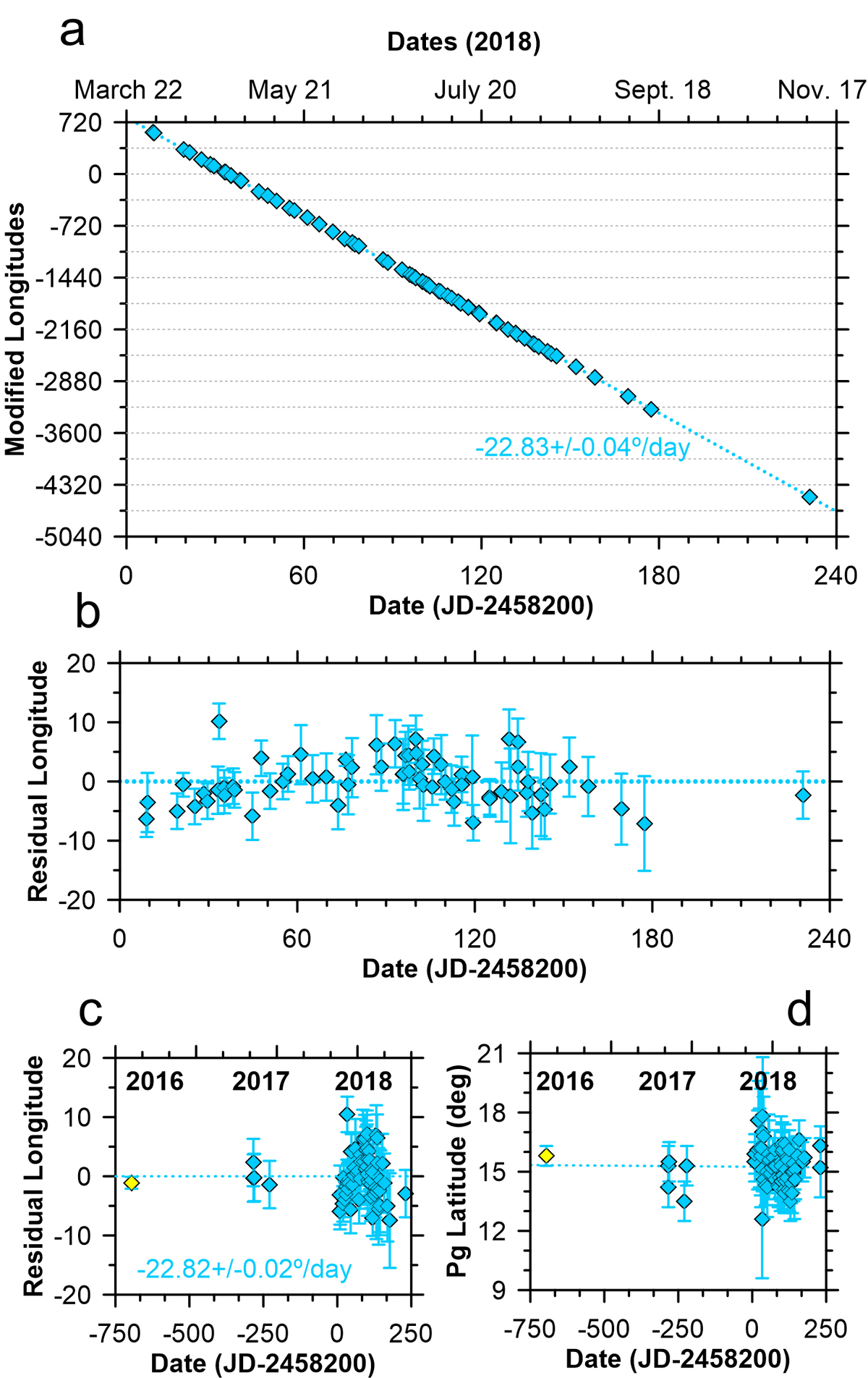}
   \caption{Tracking of the shear cloud at the northern edge of the Equatorial Zone. Panel a shows longitudinal measurements over 2018. The vertical axis is a modified system III longitude system to display data corresponding to different rotations over the planet. Grid-lines are drawn every 360$^{\circ}$ degrees. Panel b shows the residual longitude of this feature when substracting the linear drift. Panel c extends this data to measurements in 2016 and 2017.  Panel d shows the latitude of this shear-like feature. Yellow symbols are used to show the measurements on Cassini images.}
   \label{fig:Shear_Tracking}
    \end{figure}

Many observations over 2018 showed a subtle distortion in the North boundary of the Equatorial Zone at $15.2\pm0.9^{\circ}$N. In HST images this feature appears as a northwards protrusion of the Equatorial Zone and it was particularly well contrasted in blue and ultraviolet wavelengths, being also visible at the weak methane absorption band in 727 nm. A similar feature is observable in Cassini images in 727 nm in April 2016 (figure \ref{fig:figure_Cassini_Map}). The tracking of this feature was very stable over 2018: $269.8\pm1.8$ ms$^{-1}$ ($-22.83\pm0.04$  $^{\circ}$day$^{-1}$). We used this tracking to predict where this feature should have been in 2016 and 2017 and looked for it on the Cassini ISS maps in April 2018 and on selected images acquired over 2017 extending the tracking of this shear-like cloud from April 25, 2016 to to November 8, 2018. Figure \ref{fig:Shear_Examples} shows example images of this cloud in 2017 and 2018. Figure \ref{fig:Shear_Tracking} shows the tracking of this cloud feature, its residual longitudes when adjusting a linear fit to the 2018 data and the residual longitudes when extending the analyzed data to 2016.  

\subsection{Tropical latitudes: 2017-2018 (Ls=88-106)}
\begin{figure}[htp]
   \centering
   \includegraphics[width=12.50cm]{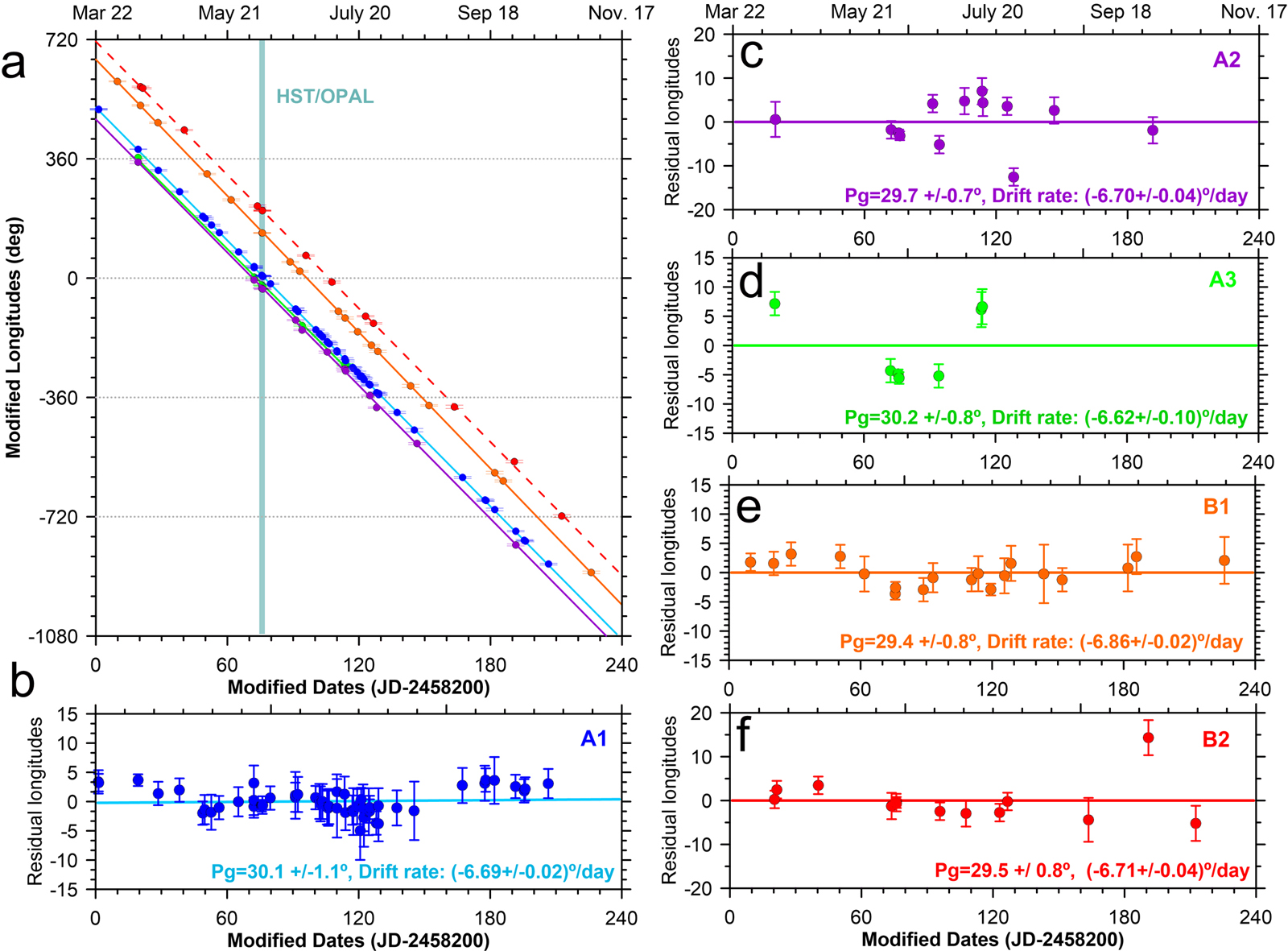}
   \caption{Tracking of features located at $\sim$30$^{\circ}$N. Panel a shows longitudinal measurements using a modified longitude system corrected for an integer number of rotations. Horizontal grid-lines are plot every 360$^{\circ}$. The date of HST observations is highlighted with a dark-blue vertical line. Features are identified by their color and linear fits to the individual features are also shown. The residual longitudes obtained from substracting the linear fit to the longitudinal positions are shown in panels b-f. Each feature is identified by a code name A1, A2, A3, B1 and B2 and can be identified in figure \ref{fig:figure_amateur_morphologies}d-f and in the HST map in Figure \ref{fig:figure_HST}. The latitude of each feature and its drift rate are indicated in panels b-f together with the uncertainties for these measurements.}
   \label{fig:Tracking_30deg2018}
    \end{figure}

Many amateur observations over 2017 and 2018 showed small nearly circular clouds at 
$\sim30^{\circ}$N with most observations showing only one feature at this latitude. Cassini images in 2016 show several cyclones at this latitude as dark patches in MT2 images accompanied by elongated clouds in the CB2 images (Figure \ref{fig:figure_Cassini_Map}).
HST observations in 2018 (Figure \ref{fig:figure_HST}) allowed to map these features close in time to the frequent amateur observations. This enabled to identify unambiguously each feature in the amateur images. Measurements of their positions over 2018 are shown in figure \ref{fig:Tracking_30deg2018}. Three features were close in longitudes (A1, A2 and A3 in the plot), and two other were nearly on the other side of the planet. Feature A1 in figures \ref{fig:figure_HST}, \ref{fig:figure_amateur_morphologies}d  and \ref{fig:Tracking_30deg2018} was easier to spot in the amateur images. Features A2 and A3 were close. The tracking of their positions show that these features separated from the linear fits around July 2018, which may have been caused by a close interaction between them. After July 20 A3 was not observed again. These cloud systems moved with a zonal speed of $72.6\pm1.0$ ms$^{-1}$ ($-6.72\pm0.04$ $^{\circ}$day$^{-1}$).

\begin{figure}[htp]
   \centering
   \includegraphics[width=12.50cm]{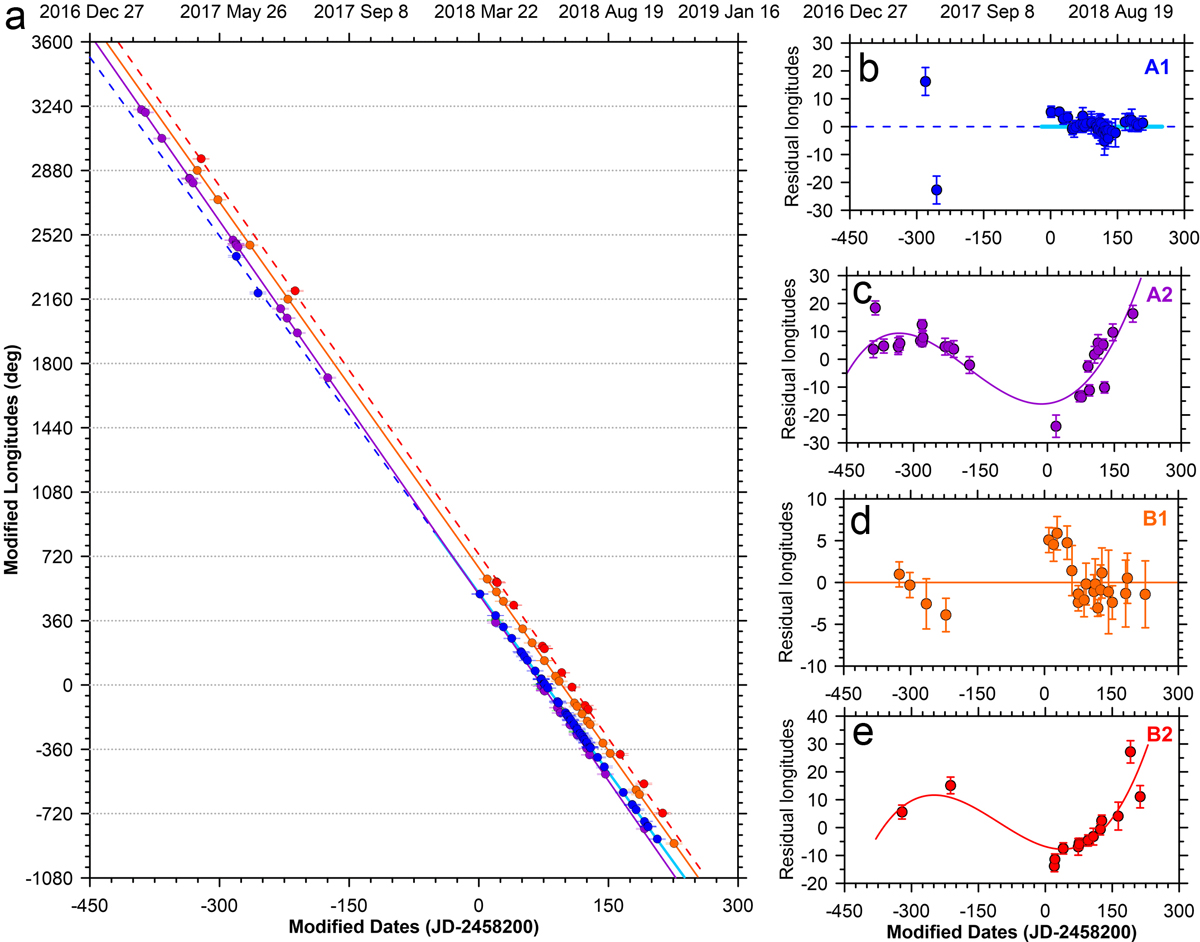}
   \caption{Extended tracking of features at approximate latitude 30$^{\circ}$N considering all the 2017 and 2018 data. Panel a shows longitudinal measurements using a modified longitude system corrected for an integer number of rotations. Horizontal grid-lines are plot every 360$^{\circ}$. Features are identified by their color and linear fits to the individual features are also shown. The residual longitudes obtained from substracting the linear fit to the longitudinal positions are shown in panels b-e. Code names are as in Figure  \ref{fig:Tracking_30deg2018}.}
   \label{fig:Tracking_30deg2017-2018}
    \end{figure}

It is possible to extend this analysis to similar cloud systems visible in 2017 (see Figures \ref{fig:figure_Pic} and \ref{fig:figure_CAHA}). Four different features were observable over 2017 in these images, although generally with a lower contrast than in 2018. Tracking these features was done considering a possible drift rate similar to those found in 2018. Results of this extended analysis are shown in figure \ref{fig:Tracking_30deg2017-2018}, which suggests that two of these cloud systems (A1 at planetographic latitude $30.1\pm1.1^{\circ}$N and A2 at planetographic latitude $29.7\pm0.7^{\circ}$N) possibly interacted very closely around January 2018 interchanging their relative positions. Feature A1 was only visible in Calar Alto observations in 2017, implying a much lower contrast that did not enable the amateur observers to detect this cloud.  We highlight the nearly round size of the cloud systems over 2017 and 2018 and their longevity implying that, most probably, they are non-convective. Cassini images in 2016 show a variety of cyclones at this latitude with different contrasts and separations among them. A one to one identification of the 2016 Cassini cyclones to the 2017 features is not possible because we lack additional data in 2016. However, we interpret the 2017 and 2018 chain of tropical features as possible cyclones.

Some images in 2018, including those acquired by HST, also showed additional cloud systems at other tropical latitudes at 19, 24 and 36$^{\circ}$N. We were not able to identify them in a number of images large enough to obtain well-defined drift rates and identifications. Three observations of the cloud at 24$^{\circ}$N (feature G1 in Figure \ref{fig:figure_HST}) fit well with a linear drift of $174\pm3.5$ ms$^{-1}$ ($-15.4\pm0.03$ $^{\circ}$ day$^{-1}$).

\subsection{Middle latitudes: Dark AV in 2011-2018 (Ls=17-104)}
\begin{figure}[htp]
   \centering
   \includegraphics[width=12.5cm]{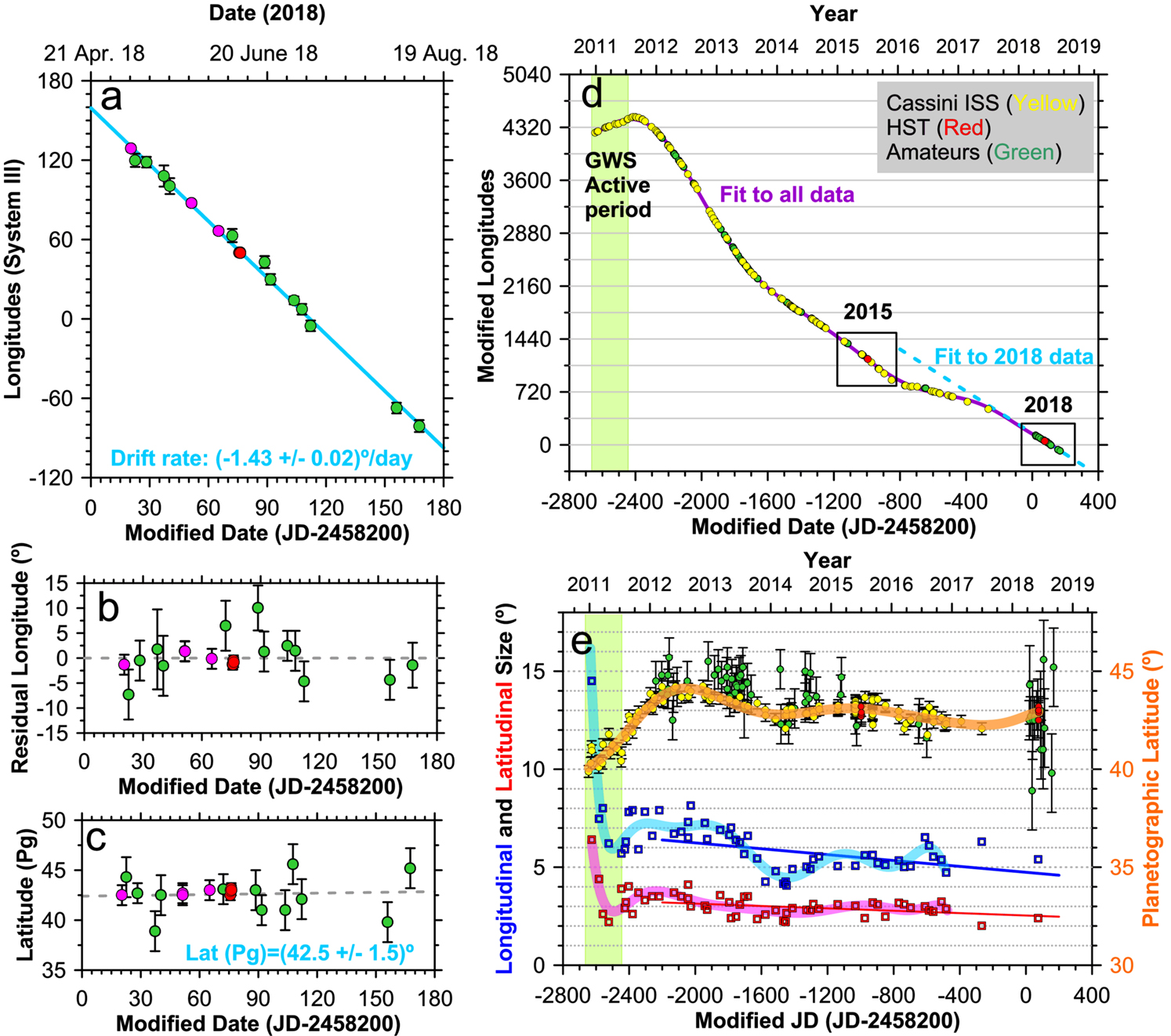}
   \caption{Mid-latitude Anticyclonic Vortex (AV). Panel a: Longitudinal positions of the vortex during 2018 on HST images (red filled circles) and exceptional amateur images showing the ring of the vortex (magenta filled circles). Green circles show measurements in images where the ring appearance of the vortex is not observable. Panel b: Residual longitudes after substracting a linear fit to the data. Panel c: Latitudinal position of the vortex. Panels d and e show the extended analysis of AV for all its life-time. Panel d: Longitudes of AV from Cassini, HST and amateur data. Panel e: size of AV in longitude (blue squares, left axis), latitude (red squares, left axis) and latitudes (all other symbols, right axis). Yellow symbols in panels d and e show Cassini ISS measurements, red symbols correspond to HST measurements, green symbols are measurements over amateur images. The green filled region shows the period of activity of the GWS (December 2010-July 2011). Thick lines correspond to polynomial fits to the data. Linear fits to the size after the initial strong changes are also shown. Longitudinal and latitudinal sizes are determined with typical errors of less than 1.0 and 0.7 degrees respectively and are not shown to highlight the fits to the data.}
   \label{fig:Dark_Vortex}
    \end{figure}

In HST images and in some exceptional amateur images, a small slightly dark vortex surrounded by a faint ring of brighter material is visible at $\sim$ 42$^{\circ}$N. The vortex is highlighted in figures \ref{fig:figure_Cassini_Map}, \ref{fig:amateur}, \ref{fig:figure_HST} and\ref{fig:figure_amateur_morphologies}g. In all amateur images in 2018 the vortex has a very small contrast with the environment resulting in a very low number of detections. We used the HST, and the three unambiguous ground-based observations of this vortex where its ring-like morphology was clearly dintiguishable, to produce a drift rate and ephemeris of its visibility over 2018. Then we searched for this vortex in images acquired by the best amateur observers over several months. We checked the images that were obtained close in time to the predicted crossing times of the feature at the central meridian of the planet. These selected images were further processed to show the faintest possible features making the vortex visible in a few of these images, although generally as a single spot without a ring. Panels a-c in Figure \ref{fig:Dark_Vortex} show the tracking of the vortex over 2018, resulting in a zonal drift of $13.4\pm0.4$ ms$^{-1}$ ($-1.43\pm0.02$ $^{\circ}$day$^{-1}$).

This vortex is the same one that formed after the Great White Spot (GWS) convective storm in 2010-2011 \citep{Fletcher2011, Sanchez-Lavega_Icarus2012, Sayanagi2013, Trammell2016}. It was called DS after Dark Spot from its visual aspect in ground based-images in \citep{Sanchez-Lavega_Icarus2012} and AV after Anticyclonic Vortex in \citet{Sayanagi2013} due to its anticyclonic circulation on Cassini ISS images. It was also visible in thermal maps of the planet as cold feature at the time of the GWS \citep{Fletcher2011}. We will name it AV in this paper. The morphology of the vortex at high resolution in Cassini ISS images can be seen in \citet{Trammell2016} and \citet{Gunnarson2018}. The AV vortex appears in images acquired in the visible as a dark elongated vortex with a faint ring of bright material around it that formed around 2013 \citep{Trammell2016} and is visible in Figure \ref{fig:figure_Cassini_Map}.

We compiled the position of this vortex in Cassini ISS, HST and amateur observations since its formation. We also measured its size in Cassini and HST images to identify the reasons for its apparent lower visibility in recent years. The size was measured in continuum band filters avoiding images close to the limb and considering the central points of the white ring of material in the north, south, east and west extremes of the vortex. The results of this analysis are shown on the right panels of figure \ref{fig:Dark_Vortex}. The vortex changed in latitude, drift rate and size over the course of eight years. The GWS storm was active from December 2010 to July 2011. During this period, AV reduced in longitude and latitude to half its original size and migrated northwards at least 2$^{\circ}$. \citet{Trammell2016} discuss the changes in size and latitude of this vortex from Cassini ISS images in the period 2011-2015 and suggest that its initial drift to the north can be explained by the beta drift effect. This northwards migration made the vortex reach latitudes where zonal winds caused a faster zonal drift. After 2013 changes were much less intense and more difficult to constrain, and after 2015 AV became less frequently observed following a sudden change in its drift rate, possibly accompanied by a small change in latitude. 

Over the last few years AV maintains a stable size and the reasons for its lower visibility on ground-based images are possibly related to a decrease of the contrast with respect to its environment in visible wavelengths. AV is currently the second longest-lived vortex observed in Saturn, after the North Polar Spot discovered in the Voyager 1 flyby in 1980 and observed until 1995 \citep{Sanchez-Lavega1997}.



\subsection{Ribbon and bright features in 2018 (Ls=99-105)}
\label{section:ribbon}

PlanetCam observations in 2017 in the 1.0-1.7 $\mu$m wavelength range and HST observations in 2018 show an undulating pattern at $\sim47^{\circ}$N. A planet-encircling wave close to this latitude was discovered in Voyager images of Saturn in 1980 and 1981 \citep{Smith1981, Smith1982}, and was called the ribbon. The ribbon was observed again in HST images in 1994-1995 with similar characteristics to the Voyager observations  \citep{Sanchez-Lavega2002}. The Cassini mission observed ribbon waves at 45-51$^{\circ}$ appearing and disappearing in different years \citep{Gunnarson2018}.  Analytical \citep{Godfrey1986}  and numerical \citep{Sayanagi2010} studies of the nature of the ribbon as observed by the Voyagers suggest the wave can be a baroclinic instability. Studies of ribbon-like waves at the time of Cassini suggest they fit better with Rossby waves with a minor baroclinic component \citep{Gunnarson2018}. 

\begin{figure}[htp]
   \centering
   \includegraphics[width=9.50cm]{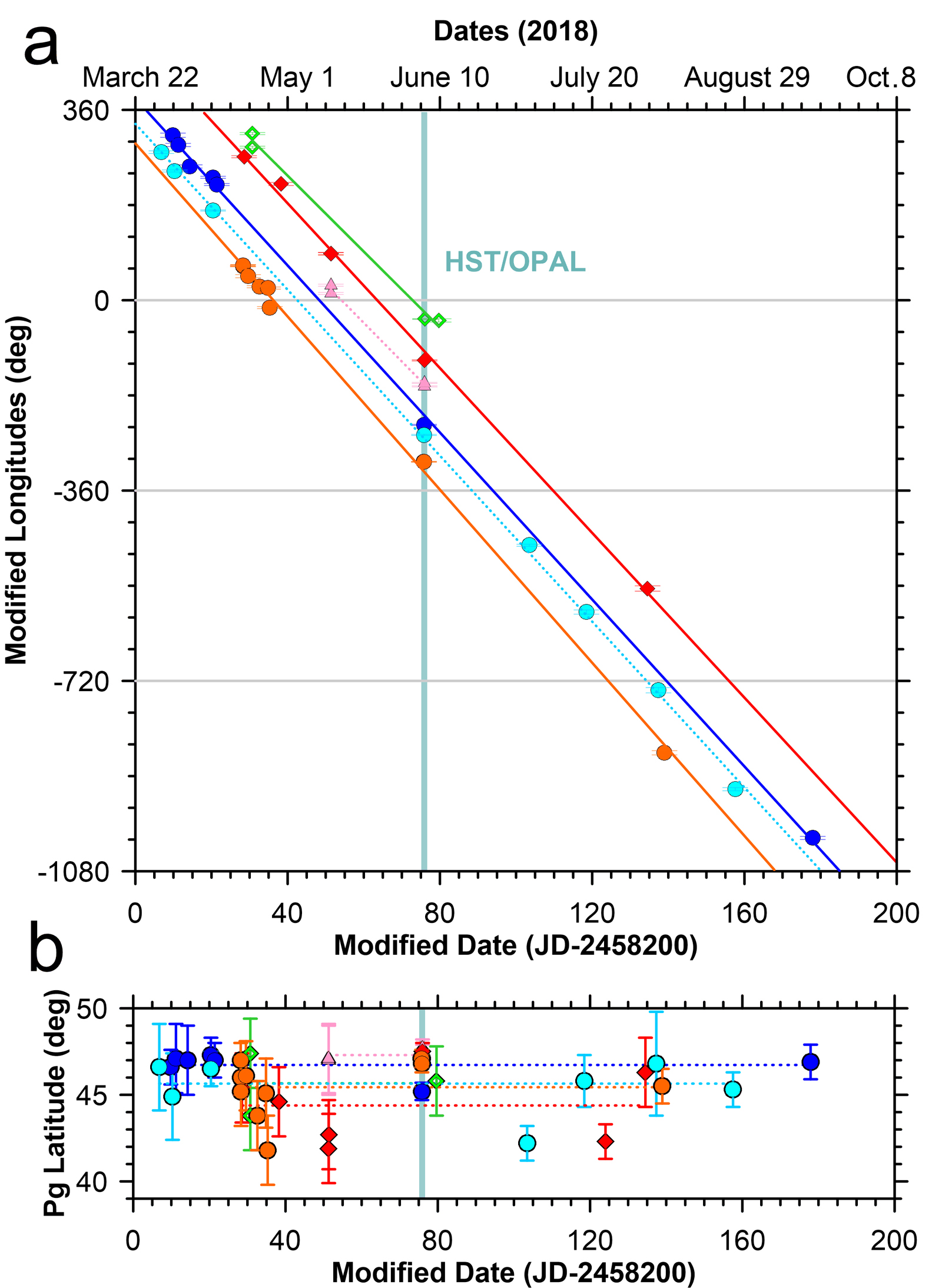}
   \caption{Tracking of bright features close in latitude to the ribbon wave. Panel a: Longitudes of different features follow similar drif rates. Mean residual longitudes to the linear fits in this plot are 10$^{\circ}$. Panel b: Latitudes of the individual features with estimated error bars. In both panels the date of HST observations is highlighted with a blue vertical box. The mean latitude of all these bright features is $45.7 \pm 1.7$.}
   \label{fig:Bright_Features}
    \end{figure}

The wavy feature in 2017 in Figure \ref{fig:figure_CAHA} constitutes the first ground-based detection of a ribbon-like wave, which has a larger amplitude and contrast in these observations when compared with the ribbon at the time of the Voyagers Images acquired over 2018 did not show the ribbon but show sometimes bright clouds at $\sim46^{\circ}$N. In spite of being generally bright (an example is given in Fig. \ref{fig:figure_amateur_morphologies}h), these clouds were very difficult to track, since they seem to appear and dissapear frequently. Their latitude corresponds to the southern limit of the ribbon wave observed on PlanetCam SWIR images, which is measured as $47.0\pm0.2^{\circ}$, and in the HST map of the planet (see Figs. \ref{fig:figure_CAHA} and \ref{fig:figure_HST}). The bright features observed at $\sim46^{\circ}$N probably correspond to locations of the main ribbon wave where the wave concentrates bright clouds or where small vortices are formed. Using as a reference the full HST map, it was possible to identify most of these features in a single drift chart by assuming a fast drift rate. Figure \ref{fig:Bright_Features} shows the drift rate of 6 possible features with similar drift rates. In HST images these features are observed as bright outcrops of the ribbon-wave and bright patches slightly south of the ribbon. The mean zonal drift for the ensemble of these features is $69\pm5$ ms$^{-1}$ ($-7.70\pm0.40$ $^{\circ}$day$^{-1}$), and their mean latitude is $45.7\pm1.0^{\circ}$N.


Finally, a very limited number of images also showed features at $51.5^{\circ}$N. The HST map shows a prominent bright feature at this latitude which is particularly well-contrasted at 502 nm (feature G2 in Fig. \ref{fig:figure_HST}). In spite of its good visibility in HST images, there are only a few instances where amateur images show weakly visible features at this latitude. Their analysis suggests a zonal drift of $26.5\pm3.5$ ms$^{-1}$ ($-3.4\pm0.3$ $^{\circ}$ day$^{-1}$).

\subsection{Subpolar latitudes: 60$^{\circ}$N vortex (2016-2018, Ls=78-106) and ACA system (2012-2018, Ls=32-102)}

In some of the best observations of 2018 a dark small spot at $\sim60^{\circ}$N is visible in the images (Fig. \ref{fig:amateur}). This spot is possibly the vortex labelled as ``A1" in the high-resolution Cassini ISS images obtained during Cassini's Grand Finale (Figure 1 in \citeauthor{Ingersoll2018} 2018) and is the subpolar single vortex described by \citet{delRio-Gaztelurrutia2018}. The vortex is also highlighted in the Cassini map in 2016 shown in Figure \ref{fig:figure_Cassini_Map}. This vortex was one of the vortices implied in a polar perturbation that developed in Saturn in May-July 2015. The perturbation was the consequence of the interaction of a system of three vortices located at 65$^{\circ}$N and named ACA after Anticyclone-Cyclone-Anticyclone with the dark vortex at 60$^{\circ}$N \citep{delRio-Gaztelurrutia2018}.

\begin{figure}[htp]
   \centering
   \includegraphics[width=7.50cm]{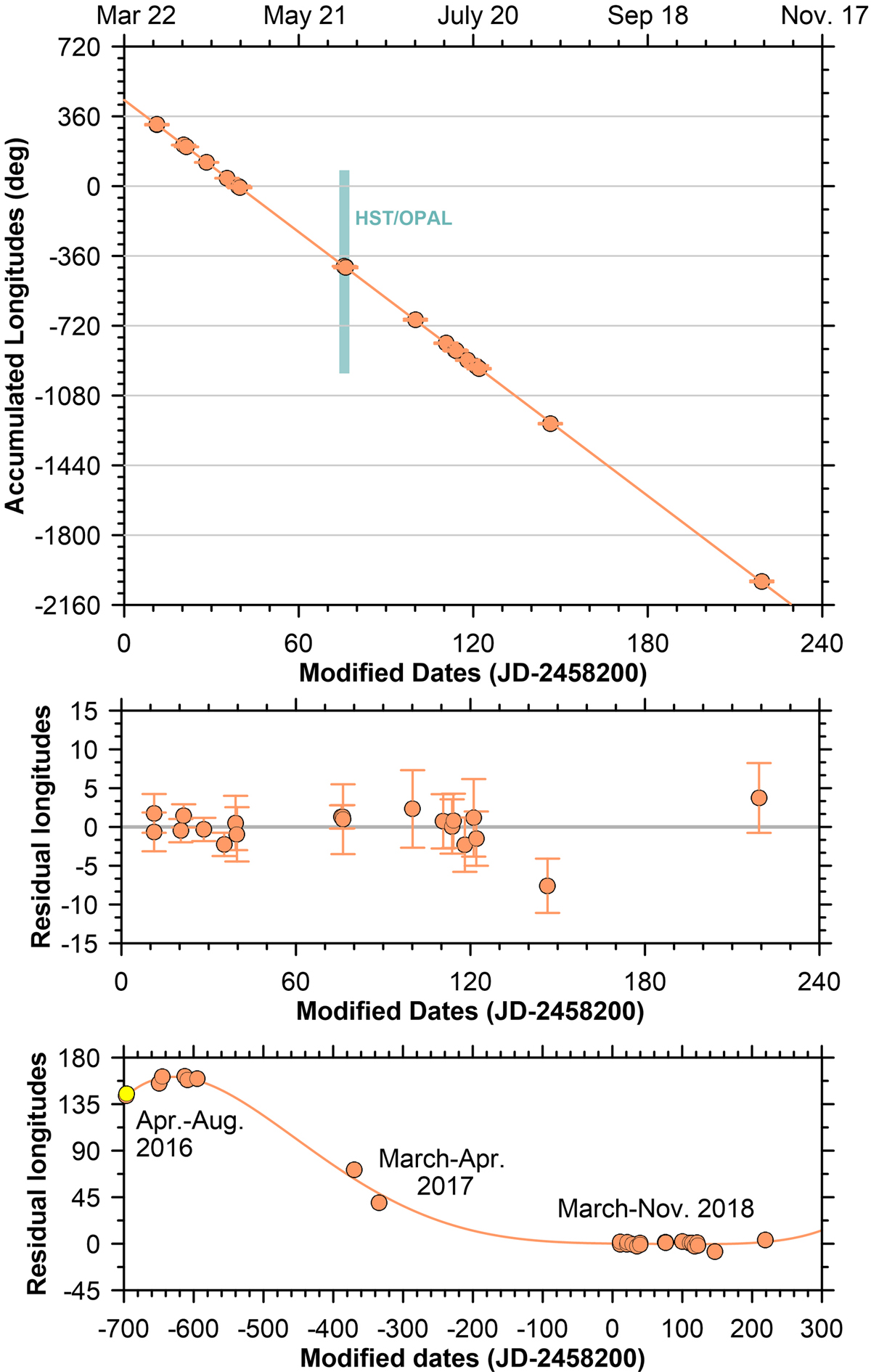}
   \caption{Tracking of the dark vortex at 60$^{\circ}$N. HST measurements are highlighted. Panel a shows the tracking of the vortex over several planetary rotations. Grid-lines are plot every $360^{\circ}$. Panel b shows differences between the longitudinal positions and a linear fit to the vortex longitudes. Panel c shows and extension of these residual longitudes considering the drift rate of 2018 and detections of this vortex in 2016 and 2017, including in Cassini images in April 2016 shown as a yellow symbol in the figure. The curved-line shows a polynomial fit to the data.}   
\label{fig:60deg_Vortex}
    \end{figure}

Our analysis of the vortex visible in 2018 is presented in Figure \ref{fig:60deg_Vortex}. The latitude of this vortex was $(60.4\pm0.9)^{\circ}$N and its zonal drift was $73.5\pm3.6$ ms$^{-1}$ $(-11.34\pm0.05^{\circ}$/day). This drift rate is stable enough to lead a search of this vortex over 2017 and 2016. We looked for this dark vortex in images acquired in 2017 and 2016 and found different images where a similar vortex was observed in March-April 2017 and April-August 2016. The bottom panel in Fig. \ref{fig:60deg_Vortex} shows the residual longitudes over 2016-2018 of all our detections of this vortex when considering only the fit to the 2018 data. The differences in the 2016, 2017 and 2018 data can be explained by a small change on the drift rate of this vortex of less than 1.6 ms$^{-1}$ (0.25$^{\circ}$day$^{-1}$).  

The ACA system at 64 $^\circ$N was first observed in Cassini ISS images in 2011 \citep{delRio-Gaztelurrutia2018}. It was a stable system of three vortices that has been observed in Cassini ISS images until 2017 (features A2, C1 and A3 in figure 1 in \citealt{Ingersoll2018}). Amateur detections of the ACA system during these years were very difficult to obtain, but the polar perturbation they produced in 2015 resulted in many detections of the vortices. In 2018 the ACA vortices were very difficult to identify and in most cases only one of the three vortices was visible as a dark small spot (the cyclone) observed at 63-65$^{\circ}$N. The different visual apperance and contrast of these vortices is related with differences in the tropospheric haze, and the higher contrast of the cyclone is caused by a lower concentration of tropospheric haze particles than its environment \citep{Sanz-Requena2019}. The cyclone was best observed in HST images using the FQ727N filter that samples the weak methane absorption band. Most probably, the low visibility of the ACA in 2018 was due to the polar storms developing at 67$^{\circ}$N, which masked the area of ACA and made it completely invisible when the polar storms grew into a full longitudinal disturbance after June 2018. The comparison of our detections of ACA in 2018 with data in 2015 and 2016 published in \citet{delRio-Gaztelurrutia2018} allowed to search for it in images obtained in 2017, where we could identify the same feature in only a few observations.

\begin{figure}[htp]
   \centering
   \includegraphics[width=12.50cm]{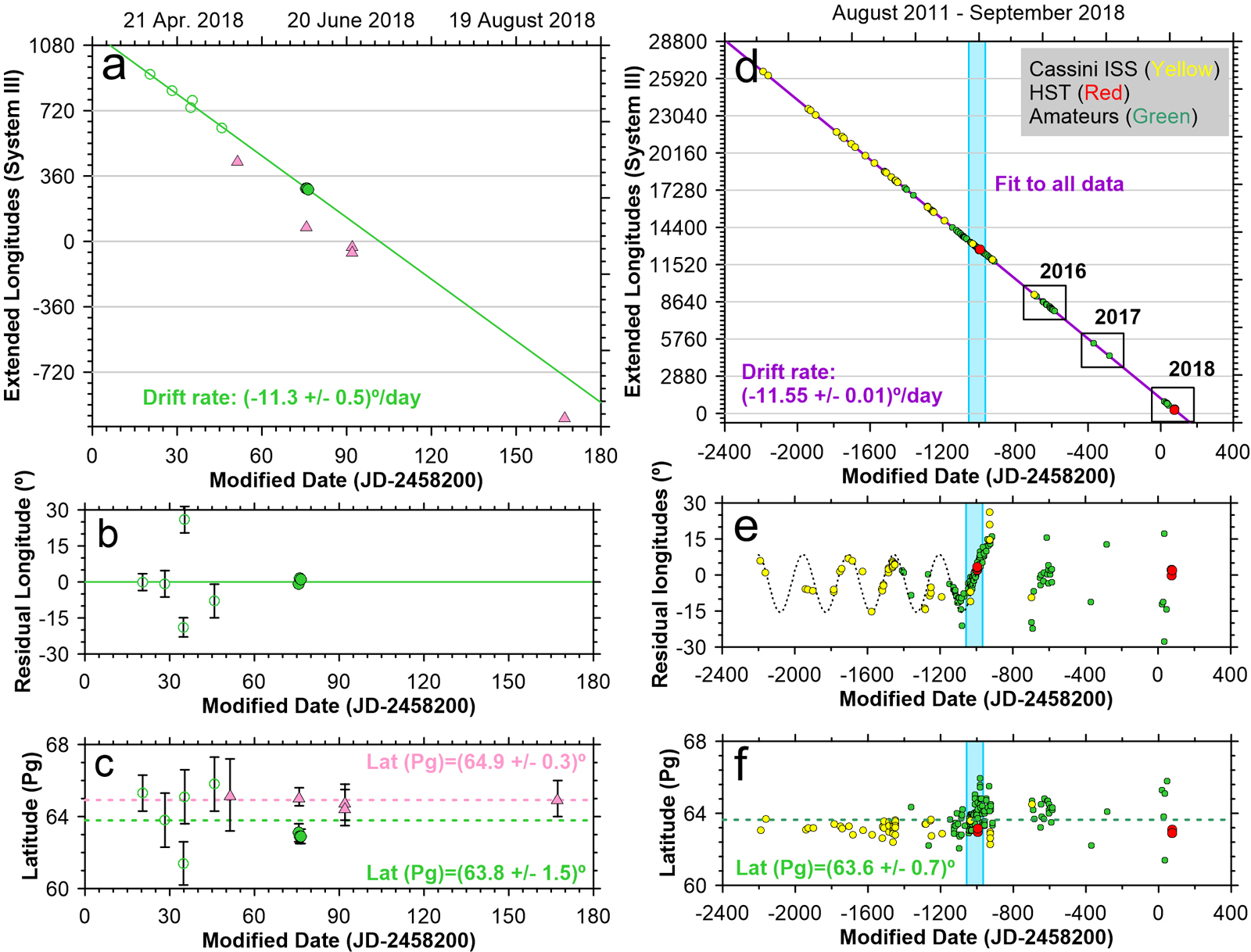}
   \caption{Tracking of the ACA system at 64$^{\circ}$N. Left panels show measurements obtained on 2018. Right panels correspond to the historical record adding measurements from \citep{delRio-Gaztelurrutia2018}. Panel a: Longitude measurements over 2018 and linear fit. Green open circles correspond to measurements on amateur images. Filled green circles to measurements on HST images. Pink triangles are measurements of similar features but that do not fit with a single linear fit. Grid-lines are shown every 360$^{\circ}$ or planetary rotation. Panel b: Residual longitudes for the ACA system after substracting the longitudinal fit. Panel c: Planetographic latitude of both features. Panel d: Longitudinal measurements of the ACA system since 2011. Yellow points are measurements on Cassini images, red dots are measurements on HST images and green dots are measurements on amateur images. A linear fit to the data is shown. The period of time shaded in blue corresponds to the activity of the disturbance discussed by \citet{delRio-Gaztelurrutia2018}. Panel e: residual longitudes over the global linear fit. A sinusoidal fit from 2011 to 2015 is shown with a dashed line. Panel f: planetographic latitudes.}   
\label{fig:ACA_Vortex}
    \end{figure}

Figure \ref{fig:ACA_Vortex} shows the tracking of this dark vortex. Left panels in this figure show the measurements in 2018. Right panels show the positions in 2016-2018 as determined in amateur images and previous measurements reported by \citet{delRio-Gaztelurrutia2018} in Cassini ISS and HST observations. The period of time when the ACA system developed a large-scale polar perturbation is shaded in blue in Figure \ref{fig:ACA_Vortex}. According to \citet{delRio-Gaztelurrutia2018}, the track of the ACA system in the period 2013-2015 exhibited oscillations in longitude with an amplitude of $\sim10^{\circ}$ and period of $\sim8.2$ months with respect to a single linear fit (see Figure \ref{fig:ACA_Vortex}e). These oscillations were accurately observed on Cassini ISS images and increased at the time of the 2015 polar perturbation reaching their maximum amplitude (nearly $20^{\circ}$). The measurements obtained in amateur images in 2016 to 2018 show a large scatter of data with respect to a linear fit that suggests that similar oscilations continued years after the perturbation. However, due to the small number of measurements in 2016 to 2018 and the errors in measurements of amateur image it is not possible to find a single oscillation with a clear period. We hypothesize that mutual interactions between the vortices in the ACA system may explain the oscillations and scatter of data in Figure \ref{fig:ACA_Vortex}e.

\subsection{Saturn's hexagon: 2015-2018 (Ls=65-106)}
Saturn's hexagon, a wave feature that distorts Saturn's Eastward zonal jet at 75.8$^{\circ}$N, was discovered on images acquired by the Voyagers in 1980 and 1981 \citep{Godfrey1988}. It was later observed in 1990-1995 with HST and in some ground-based images \citep{Sanchez-Lavega1993, Caldwell1993}. In Voyager and early HST images, a large anticyclone outside the hexagon, the so-called north polar spot (NPS), drifted in longitude at a similar rate as the hexagon and was proposed to imprint a perturbation on the polar jet that could cause the hexagon \citep{Allison1990,Godfrey1990,Sanchez-Lavega1997}.

Cassini observed the hexagon first in thermal-IR images in 2007 during winter in Saturn's north hemisphere demostrating that the hexagon was present in Saturn's atmosphere without solar insolation and hinting to a dynamical origin deeper than the main cloud level \citep{Fletcher2008, Baines2009}. The hexagon was later observed in Cassini ISS images in visible light since 2012 until the end of the mission \citep{Sanchez-Lavega2014, Antunano2015, Sayanagi2018} and eventually also manifested in the stratosphere in 2014 \citep{Fletcher2018NatCo}. The NPS that had been observed in the 1980s and 1990s was not observed in any of these thermal infrared or visible images. 

Although the hexagon lies at a fast Eastward jet, the positions of the vertices that define its shape are almost steady with respect to Saturn's system III longitudes. \citet{Sanchez-Lavega2014} ran an analysis of the hexagon's drift rate in time comparing data obtained from the Voyagers in 1980-1981, HST and ground-based observations in 1990-1991 and Cassini ISS images in 2012-2014 concluding that the hexagon should have a deep structure responsible of its stability over such a long period of time. Amateur observations since 2015 allow to extend this study. 

\begin{figure}[htp]
   \centering
   \includegraphics[width=10.00cm]{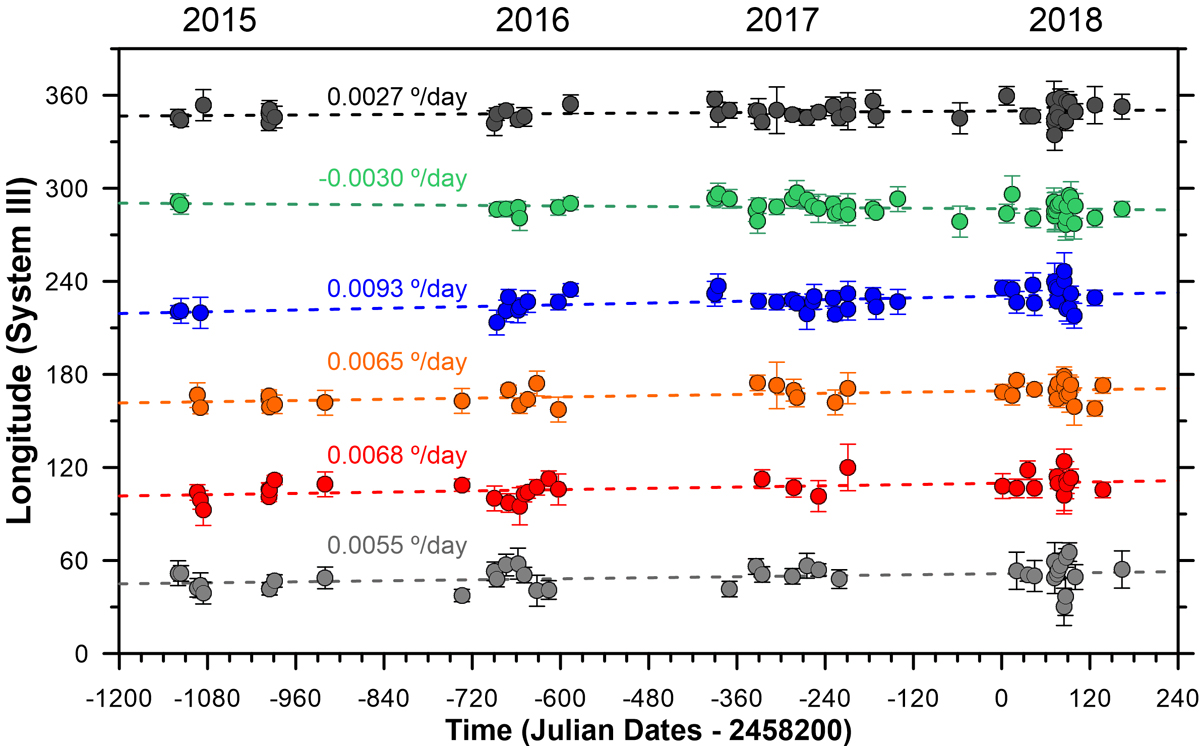}
   \caption{Positions of the vertices of Saturn's hexagon on amateur images over 2015-2018. Each color represent the position of a different vertex of the hexagon. Measurements in HST, Calar Alto and Pic du Midi images are also incorporated. Error bars are estimations based on the difficulties in identifying a unique position for the vertex and maximize for vertices far from the central meridian.}   
\label{fig:Hexagon_new_data}
    \end{figure}

We obtained measurements of the vertices of the hexagon in selected amateur observations in 2015-2018. The visibility of the hexagon in 2015 was limited, resulting in a lower number of measurements. More precise measurements of the vertices were obtained in 2016-2018. Figure \ref{fig:Hexagon_new_data} shows these measurements. Drift rates for each vertex in this period lie in the range from $-3.0 \times 10^{-3}$ $^{\circ}$day$^{-1}$ to $9.3 \times 10^{-3}$ $^{\circ}$day$^{-1}$ with a mean value of $(4.6\pm4.3)\times10^{-3}$ $^{\circ}$day$^{-1}$  ($0.012\pm0.011$ ms$^{-1}$) that makes the hexagon almost stationary with respect to system III longitudes.

These numbers are comparable, but slightly different, to  measurements in Cassini ISS given by \citep{Sanchez-Lavega2014} and in Cassini CIRS images in \citep{Fletcher2018NatCo}. We will analyse the significancy of these differences in the discussion.

\section{Discussion: Drift rates and winds}

\begin{figure}[htp]
   \centering
   \includegraphics[width=12.50cm]{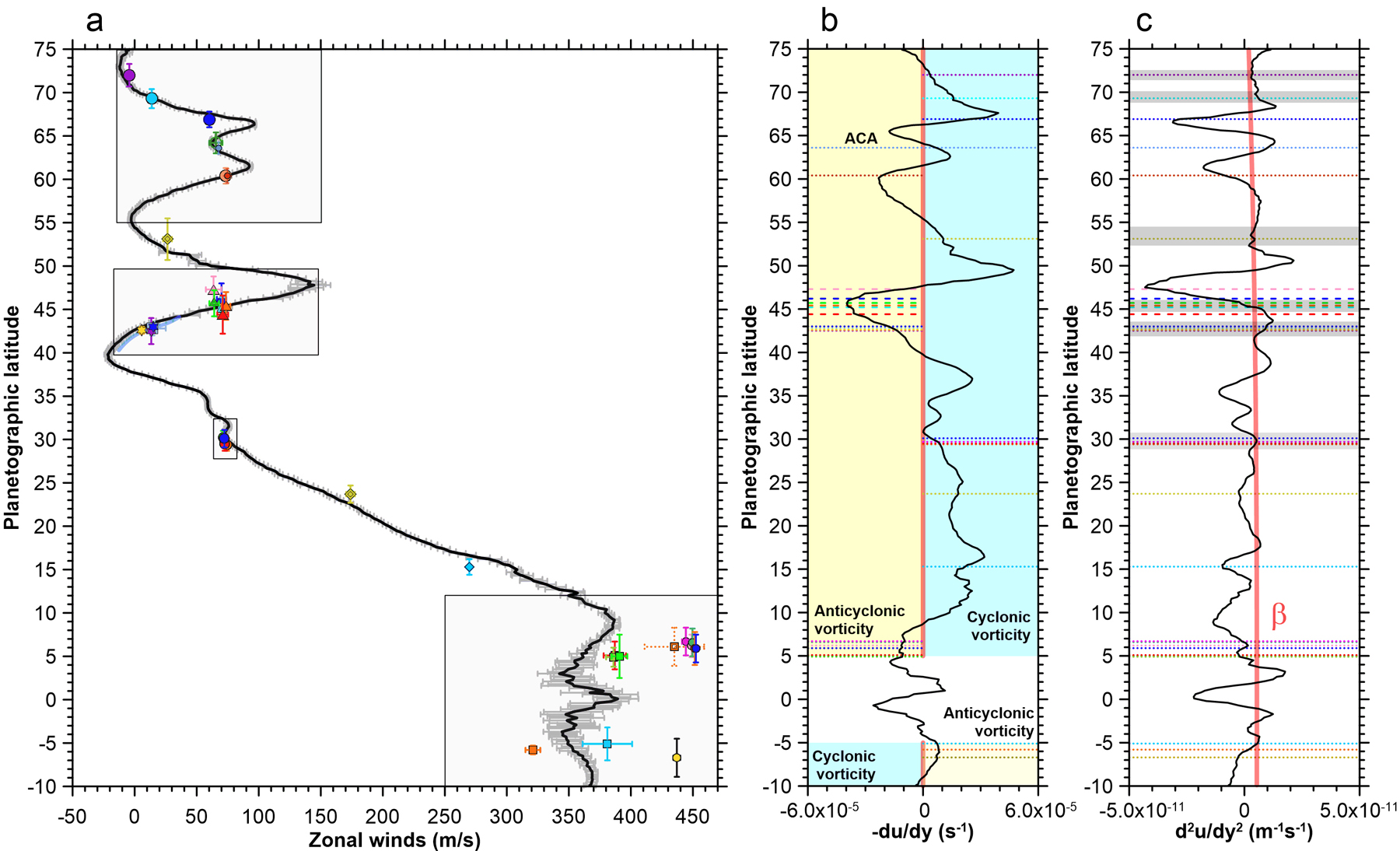}
   \caption{Saturn's zonal winds and atmospheric features. a) Zonal winds from Cassini ISS images in\citet{Garcia-Melendo2011} (black line with error bars) compared with drift rates of atmospheric features in this work (symbols). Symbols and colors are as in previous figures and features at latitudes 67$^{\circ}$N, 69$^{\circ}$N and 72$^{\circ}$N are taken from \citep{Sanchez-Lavega2019}.  The horizontal grey dotted line shows the latitude of the Shear-like feature at 15$^{\circ}$N which is located exactly at the location of a small local variation of the zonal winds. Boxes indicate regions zoomed in figures \ref{fig:Winds_Specific_Equator} and \ref{fig:Winds_Specific}. b: Relative vorticity indicating regions of cyclonic and anticyclonic vorticity. c: Meridional gradient of the relative vorticity compared with the $\beta$ parameter. In b and c horizontal lines indicate the latitudinal location of the clouds features studied. The first and second meridional derivatives of the zonal winds shown in these panels depend on the latitudinal step used to compute the derivatives. Values in panels b and c have been smoothed with an scale of 1$^{\circ}$ with respect to the 0.2$^{\circ}$ meridional resolution of the zonal winds.}   
\label{fig:Winds_General}
    \end{figure}

The drift rates of the cloud features described above are compared with Saturn's zonal winds \citep{Garcia-Melendo2011} in figures \ref{fig:Winds_General}, \ref{fig:Winds_Specific_Equator}, and \ref{fig:Winds_Specific}. For completeness, these figures also incorporate the drift rates of the main convective storms in the polar area over 2018 and named White Spot 1 (WS1), White Spot 2 (WS2) and White Spot 3 (WS3) in \citet{Sanchez-Lavega2019} and found at latitudes 67$^{\circ}$N, 69$^{\circ}$N and 72$^{\circ}$N. 

Figure \ref{fig:Winds_General}a shows a global view of the winds at cloud level and the position and zonal spped of the atmospheric details tracked in this work. Panel b shows the vorticity associated to the zonal winds ($-du/dy$) and allows to identify which features are most probably cyclones and anticyclones by the sign of the relative vorticity of the zonal winds (note that the ACA system is located in a region of nearly null relative vorticity allowing the close latitudinal position of the cyclone and the anticyclones in this sytem). Panel c shows the curvature of the zonal winds ($u_{yy}=d^2u/dy^2$), or meridional gradient of vorticity, compared with the $\beta$ parameter ($\beta=df/dy=2\Omega cos\phi/R_s$, where $f$ is the Coriolis parameter, $\Omega$ is the system III rotation angular velocity, $\phi$ is planetographic latitude and $R_s$ is Saturn radius at a given latitude). Latitudes in which the gradient of absolute vorticity $(\beta-u_{yy})$ change sign satisfy a necessary, but not sufficient, condition for barotropic instabilities to develop known as the Rayleigh-Kuo criterion. \cite{Read2009} examined Voyager zonal wind profiles and Cassini CIRS maps of temperature of Saturn to map potential vorticity and identify regions of instability in the planet. In that work many of the eddy structures and planetary waves are located at latitudes where  $(\beta-u_{yy})$ changes sign. We show in Figure \ref{fig:Winds_General}c the latitudes where long-lived features were found. Those where $(\beta-u_{yy})$ changes sign are highlighted and include the chain of cyclones at 30$^{\circ}$N, the AV vortex in its southern limit, the bright features associated to the ribbon, the feature at 53$^{\circ}$N and the White Spot 3 formed after the onset of the bright convective polar storms discussed in \citep{Sanchez-Lavega2019}.

Features in the Equatorial Zone are left out of this analysis due to the minor role of Coriolis forces in the Equator and we remark that these cloud systems are the only ones that present large differences in their velocities with the Cassini zonal winds.

\subsection{Winds and atmospheric features in the EZ}

The Equatorial Zone (EZ) is particularly complex, as it is a region known to have temporal changes in the wind field at cloud level and in its vertical wind shear \citep{Sanchez-Lavega2003, Porco2005} over most of the troposphere, as well as quasi-periodic thermal oscillations in the stratosphere that occur both in altitude and in time \citep{Fouchet2008,  Orton2008, Guerlet2018}. The quasi-periodic oscillations of the stratosphere where disrupted with the development of the 2010-2011 Great White Spot \citep{Fletcher2017} further complicating the dynamics of the EZ. 

\begin{figure}[htp]
   \centering
   \includegraphics[width=10.00cm]{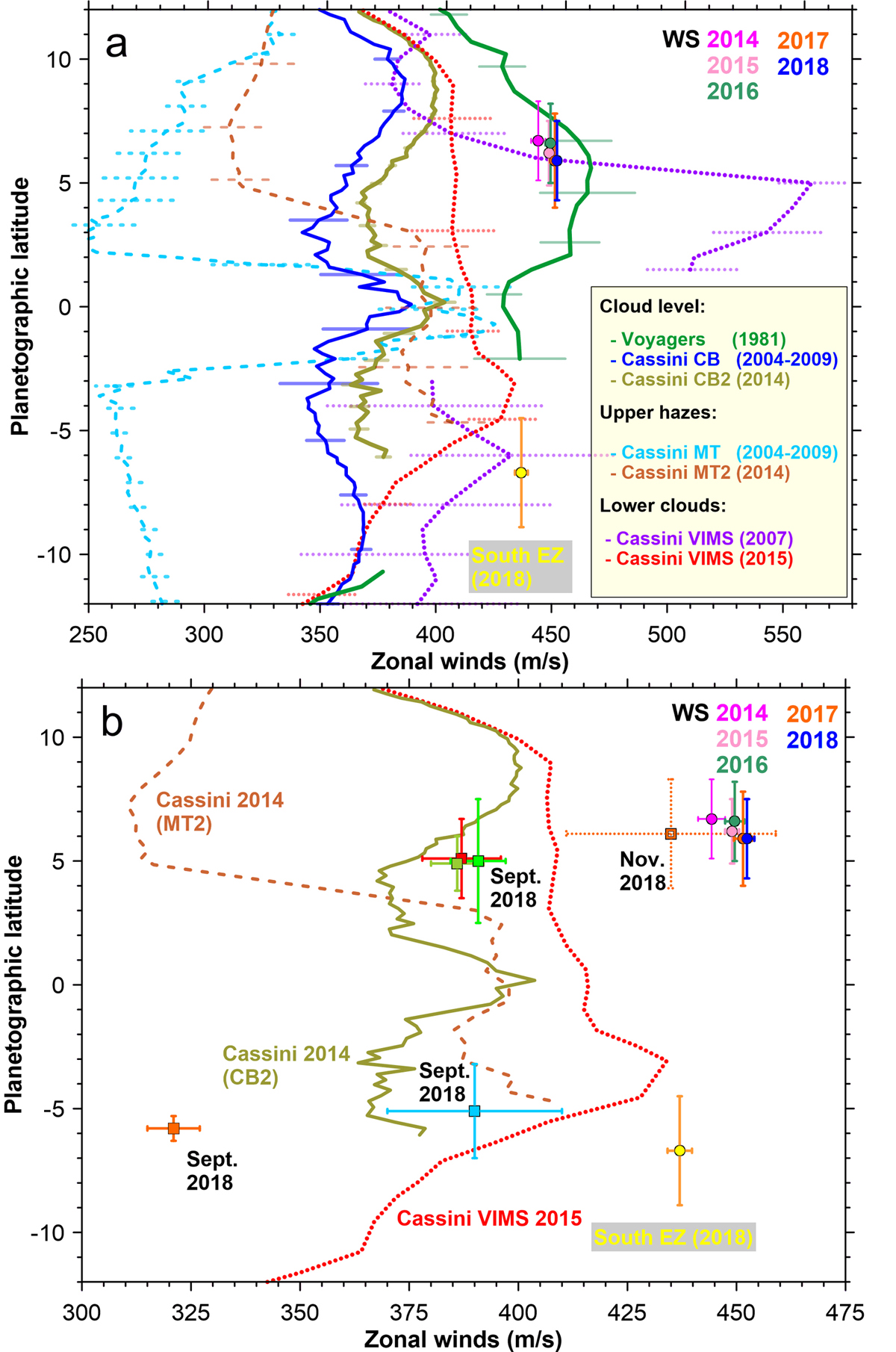}
   \caption{Saturn's zonal winds from spacecraft images compared with velocities of the features tracked on ground-based images. a: Zonal winds in a variety of wavelenghts and years including 
Cassini ISS images in continuum band filters (solid lines), methane band filters (dashed lines), VIMS thermal images (dotted lines) and the Voyager wind profile (solid dark green line). The bright White Spot in the North Equatorial Zone in 2014-2018 and the bright White Spot in the South Equatorial Zone in 2018 are shown with symbols as in previous figures. b: Most recent zonal winds from Cassini obtained in 2014 (continuum band and methane band) and VIMS data in 2015 compared with all equatorial features tracked in this work.}   
\label{fig:Winds_Specific_Equator}
    \end{figure}

\begin{figure}[htp]
   \centering
   \includegraphics[width=5.30cm]{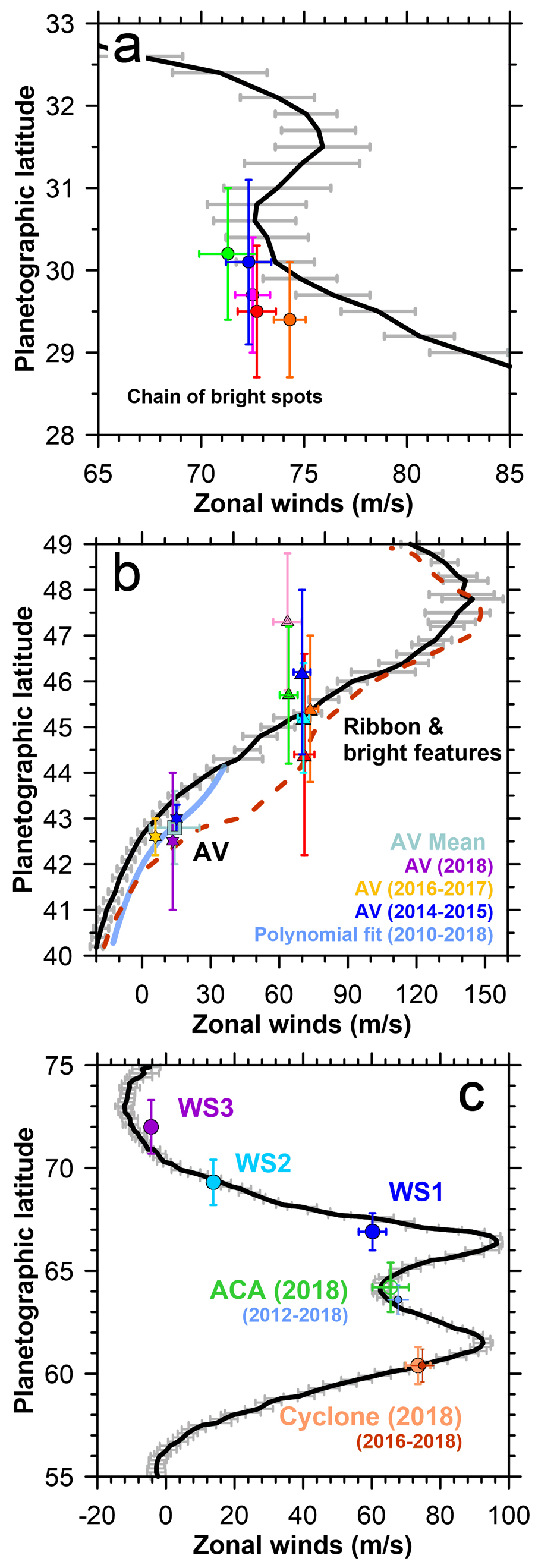}
   \caption{Saturn's zonal winds on Cassini ISS images from \citet{Garcia-Melendo2011} (black line with error bars) compared with drift rates of atmospheric features in this work (symbols). Symbols and colors are as in previous figures. Panel a: tropical latitudes; panel b: mid-latitudes; panel c: polar latitudes. Labels in the figure indicate the period of time of the data analyzed for each feature. For the AV vortex different symbols identify particular epochs. A blue thick line shows the result of transforming a polynomial fit to the longitudes and latitudes of the vortex as a function of time into zonal winds.  The dashed dark red line in panel b is the zonal winds in August 2011  \citep{Sayanagi2013}.  Features WS1, WS2 and WS3 are the two main convective storms and an additional feature described in  \citep{Sanchez-Lavega2019}.}   
\label{fig:Winds_Specific}
    \end{figure}

Winds and motions of equatorial features are presented in detail in figure \ref{fig:Winds_Specific_Equator} in two panels. The first panel shows the temporal evolution of the zonal winds at different altitudes compared with the main features observed in ground-based images from 2014 to 2018. The second panel shows all features tracked in ground-based images compared only with the most recent Cassini wind profiles obtained in 2014 and 2015. Data in this figure comes from Voyager images in 1980 and 1981 \citep{Sanchez-Lavega2000}, Cassini ISS data obtained in the years 2004-2009 \citep{Garcia-Melendo2010, Garcia-Melendo2011} and in 2014 \citep{Sanchez-Lavega_NatCom2016}, and Cassini VIMS images in 2007 \citep{Choi2009}, and 2015 \citep{Studwell2018}.  The Cassini ISS data corresponds to two different layers sensed in CB filters sensitive to the top of the main cloud layer and MT filters sensitive to the upper hazes in the lower stratosphere at $\sim60$ mbar. Motions at both layers are very different in the equator for the same years and the narrow and intense jet in the upper hazes and its temporal variations could be related to the equatorial thermal oscillations \citep{Fouchet2008, Orton2008, Guerlet2018}. The Cassini VIMS data senses deeper clouds at roughly 2 bar, where temporal changes are not expected. The apparent differences between both Cassini VIMS wind profiles might be a consequence of the limited spatial resolution and latitudinal coverage of the 2007 VIMS data, with significant scatter in the wind measurements. Temporal differences between Voyager and Cassini winds have been studied with radiative transfer calculations that show that dynamical effects are required to explain temporal and vertical differences between the winds obtained from Voyager color images and Cassini ISS images in CB filters \citep{Perez-Hoyos2006b, Sanchez-Lavega_NatCom2016}. 


The bright white spot (WS) observed from 2014 to 2018 roughly matches the VIMS zonal winds obtained in 2007 and the Voyager zonal winds at the cloud level, although the VIMS zonal winds in 2007 were highly uncertain in the equator due to the limited spatial resolution of the observations and VIMS zonal winds in 2015 had relatively larger measurements errors estimated to be on the order of 18 ms$^{-1}$. Measurements of the WS in different years are very similar and confirm previous results for this bright feature in 2014-2015 \citep{Sanchez-Lavega_NatCom2016}, where the interpretation of the data was that the WS has a deep root at the cloud layer observed in VIMS images. This interpretation was based on two factors: (i) radiative transfer modeling of the 2015 HST data suggested that the WS cloud-top was located at $1.4\pm0.7$ bar, with the rest of the equatorial zone moving at smaller velocities compatible with Cassini ISS zonal winds and located at 400-700 mbar. (ii) The coincidence in the zonal speed of the WS with the VIMS zonal winds from 2007 data. However, the VIMS zonal winds in 2015 published later \citep{Studwell2018} suggest that the WS moves sligthly faster than the deep winds, complicating this interpretation. The new WS ground-based data in 2017-2018 are more precise and show a sustained  fast feature  with a continuous acceleration of its zonal speed from $444.3\pm3.1$ ms$^{-1}$ in 2014 to $452.4\pm1.7$ ms$^{-1}$ in 2018 with an overall acceleration of 1.9 ms$^{-1}$ per year.  This can be compared with an acceleration of the Cassini zonal winds in CB filters of 10-20 ms$^{-1}$ from 2004-2009 to 2014, which in turn is comparable to the $5-10$ ms$^{-1}$ acceleration of equatorial winds in the south equatorial zone reported by  \citet{Li2011}  also in CB filters with stronger changes in the methane band filters.

The features that separated from WS in early August 2018 moved at a slower rate over September 2018. Their velocities are compatible with the most recent measurements of cloud motions at the visible cloud layer with Cassini ISS images in 2014. These elongated features were possibly affected by the vertical wind shear of the atmosphere when they detached from the WS. A short-lived feature observed in November 2018 moved at the same rate as the WS.

The main bright White Spot in the South EZ visible over the first half of 2018 also moves at a fast wind speed that is compatible with deep winds from the VIMS zonal wind profile and not compatible with the most recent winds obtained at the cloud level from Cassini. Its morphology is very similar to the morphology of the WS and is located at a symmetric latitude with respect to the Equator. We consider that both features are deep-rooted cloud systems that move with velocities characteristics of the 2 bar level with their top clouds observed at the visible cloud layer. Additional features in the SEZ that appeared over September 2018 moved at slower velocities. One of these cloud systems moved with the same wind speed as those measured at cloud level on Cassini ISS images. The second feature moved much more slowly and its zonal speed is intermediate between those of the winds at the main cloud and the upper hazes on Cassini ISS images. The three different velocities at the SEZ correspond to the same epoch and constitute a strong evidence of an intense vertical wind shear in the SEZ.

Finally, the shear-like feature in the Northern edge of the EZ also moves at a different speed than the zonal winds, moving $\sim$35 ms$^{-1}$ slower than the zonal winds obtained by Cassini at the same latitude. This feature is located at a very particular latitude in the limit of the hazes that cover the EZ. This is also a region where the zonal winds have a kink, centered at a minimum of relative vorticity and a local maximum of the meridional gradient of vorticity. These properties may indicate  a possible wave instability. In that case, the difference of 35 ms$^{-1}$ in the zonal wind of the shear-like cloud and the zonal winds may represent a phase speed of the wave,  which has remained largely stable at least since 2016.

\subsection{Winds and atmospheric features in the tropics, mid-latitudes and subpolar latitudes}

Figure \ref{fig:Winds_Specific} presents zooms over additional  features tracked in this work. The tropical features moved with a similar velocity to the zonal winds at these latitudes, with differences on the order of 1-4 ms$^{-1}$. The chain of details at slightly different latitudes seem to follow the meridional wind shear of the zonal winds,  they should be cyclonic in nature and are located at a latitude that satisfies the Rayleigh-Kuo neccessary condition for barotropic instability .

In the mid-latitudes there are two types of features visible in the amateur images: ({\em i}) The large and long-lived AV vortex, and ({\em ii}), the bright and short-lived features probably associated to the ribbon. 

The latitudinal motions of AV since it formed in 2011 result in motions with different zonal winds in different periods of time. Figure \ref{fig:Winds_Specific} shows mean motions for different periods of time and the result of a fit to all data. This fit was obtained by deriving in time the fit to the longitudinal behavior, transforming the result into velocities in ms$^{-1}$, representing zonal winds as function of latitude, and computing a third-order polynomial fit to the result. This final fit closely matches Cassini zonal winds in 2004-2009 fully reproducing the meridional variations of the winds but resulting in drift rates of the vortex $5$ ms$^{-1}$ faster than the zonal winds.  The faster velocity of AV with respect to Cassini winds from 2004-2009 might be a consequence  of small variations in the zonal winds occurring after the 2010-2011 GWS, resulting in slightly faster zonal winds after the storm from 40-46$^{\circ}$N. \citet{Sayanagi2013}  studied changes in the zonal winds in Saturn associated to the 2010-2011 GWS before and right after the end of the convective activity. AV's drift rate over 2010-2018 is intermediate between the Cassini winds in 2004-2009 and the winds measured in August 2011, right after the end of the convective activity, and measured over a highly perturbed atmosphere. 

The bright features close in latitude to the ribbon were more difficult to identify and had large errors associated to their drift rates and main latitude. Most of these cloud systems moved at velocities compatible with the zonal winds from Cassini images. 
A statistical analysis of the differences between the zonal winds and the bright features implies a difference of $-15\pm28$ ms$^{-1}$. 
These differences have error bars that are too large to distinguish if the bright features are advected by the wind or are a manifestation of a phase speed of the ribbon.

 Subpolar vortices in 2018 like the ACA and the 60$^{\circ}$N vortex follow closely the zonal winds. When using longitudinal positions over previous years, their drift rate matches the Cassini zonal winds within 2 ms$^{-1}$. 

Finally, we complete the study of details in the polar latitudes with the measurements of the mean motions of the convective polar storms visible over most of 2018  (WS1, WS2 and WS3 in the HST map in Figure  \ref{fig:figure_HST}) and described in detail in \citet{Sanchez-Lavega2019}. The motions of the polar storms roughly followed the zonal wind profiles, but differences between the drift rate of the first storm (WS1) and the zonal winds are significant and can not be only attributed to the uncertainties in the measurements.

\section{Discussion: The North Polar Hexagon drift rate in time}
Saturn's hexagon is known to have a very small drift rate with respect to Saturn's System III longitude system. This drift rate, when interpreted as a phase speed with respect to the environment wind, is consistent with a Rossby wave \citep{Allison1990,Sanchez-Lavega2014}. Other mechanisms have been examined in the literature, like barotropic linear instabilities \citep{Barbosa-Aguiar2010}, or direct forcing of the hexagon by nearby vortices \citep{Morales-Juberias2011}. However, these alternatives predict regular patterns of vortices that are not observed in Cassini images, or phase speeds for the hexagon that are far from those observed. More recent simulations showed that small perturbations to an eastward jet can grow and form a meandering jet when the jet latitude, amplitude, curvature  and vertical structure are appropiate \citep{Morales-Juberias2015}. These simulations do not require nearby vortices and they succesfull reproduce the hexagon's characteristics by imposing conditions to the zonal jet at deep levels (at least 2 bar with a constant behavior in depth until 10 bar), well bellow the upper atmosphere, which is sensitive to seasonal effects.

\begin{sidewaystable} %
\centering
\begin{tabular}{l l c r}\hline
Time period         &   Data               &  Drift rate ($^{\circ}$day$^{-1}$) & Reference\\
\hline
1980-1981           &  Voyager 1 and 2                 &  $-0.0602  \pm 0.014$   & \citep{Sanchez-Lavega2014}\\
1990-1991           &  HST and Pic du Midi            &  $-0.0010  \pm 0.007$   & \citep{Sanchez-Lavega2014}\\
1990-1991           &  HST, Pic du Midi and HST     &  $-0.036  \pm 0.0021$  & This work\\
2007-2011           &  Cassini CIRS and ISS           &  $+0.0131 \pm 0.004$   & This work\\
2008-2014           &  Cassini ISS                         &  $+0.0129 \pm 0.002$   & \citep{Sanchez-Lavega2014}\\
2014-2017           &  Cassini CIRS                       &  $+0.099   \pm 0.013$  & \citep{Fletcher2018NatCo}\\
2014-2017           &  Cassini CIRS (stratosphere)   &  $+0.076   \pm 0.019$  & \citep{Fletcher2018NatCo}\\
2015-2018           &  Ground-based                     &  $+0.0046 \pm 0.0043$ & This work\\
\hline 
\end{tabular}
\caption{Drift rates of the hexagon with respect to System III in different periods of time.}
\label{tab:Hexagon_drifts}
\end{sidewaystable}
Table \ref{tab:Hexagon_drifts} shows measurements of the hexagon's drift rate as a function of time in different years from different sources, including our current measurements on amateur images from 2015-2018. We also incorporate here a slight improvement of the 1990-1991 drift rate from using an additional HST image obtained in July 1991 and where the hexagon is easily observable \citep{Westphal1991}. Further measurements of Saturn's hexagon drift rate are given from comparing Cassini CIRS images of the hexagon obtained in March 2007 with Cassini ISS images in  2008-2011. Our measurements of Cassini CIRS images are based on the synthetic map presented by \citet{Fletcher2008} and our measurements of the vertices of the hexagon in 2008-2011 come from \citet{Sanchez-Lavega2014}. We also list here additional measurements of the hexagon's drift rate from Cassini CIRS images in 2014-2017 analyzed by \citet{Fletcher2018NatCo}. The drift rates of the hexagon in the 1980s and 1990s can be compared with measurements of the drift rate of the NPS. These were $-0.061$ $^{\circ}$day$^{-1}$ in 1980-1981 and $-0.030$ $^{\circ}$day$^{-1}$ in 1990-1995 \citep{Caldwell1993, Sanchez-Lavega1997} and are within the error bars of the measurements.

   \begin{figure}[htp]
   \centering
   \includegraphics[width=10.50cm]{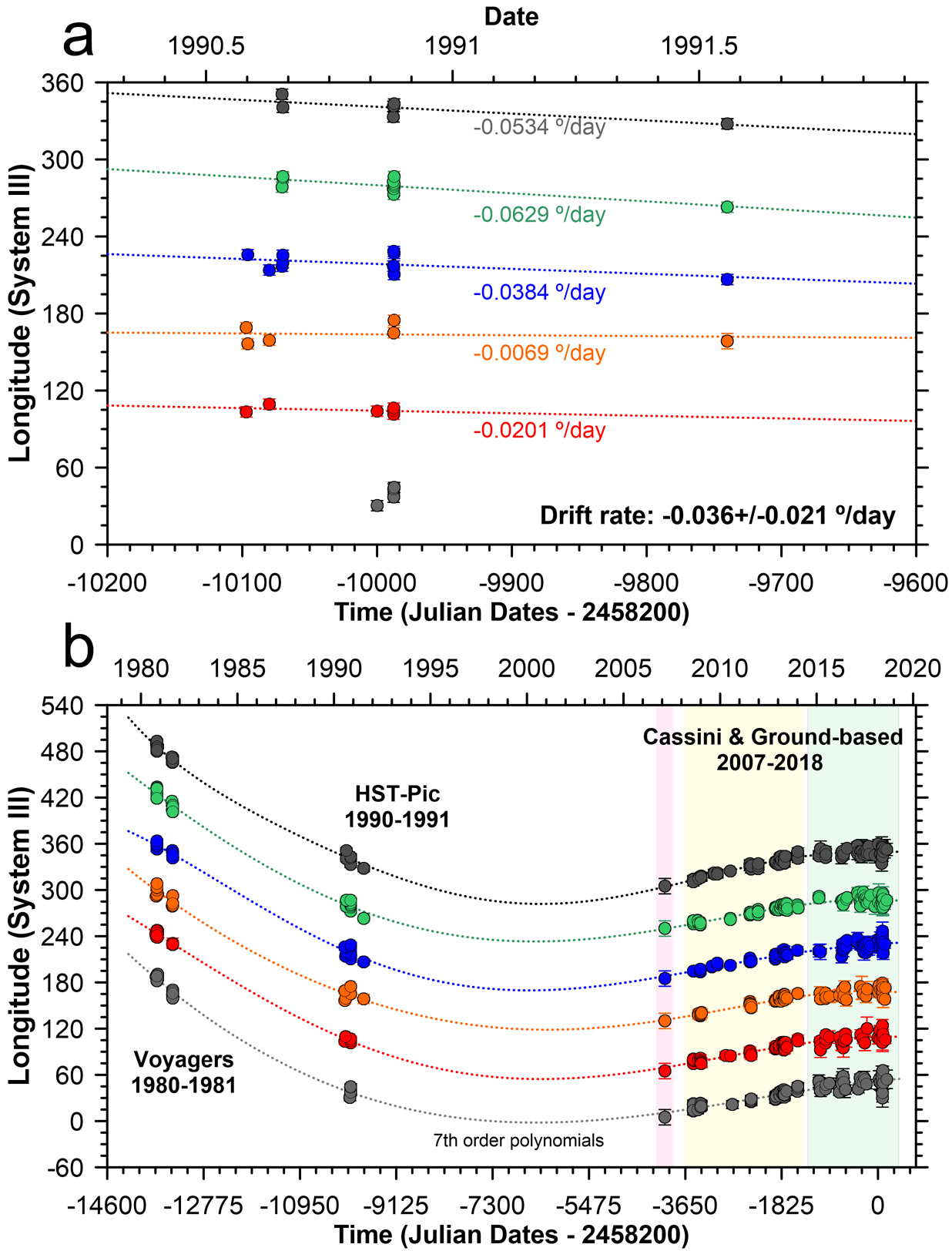}
   \caption{Positions of the hexagon vertices with respect to System III. (a) HST and Pic du Midi data from \citep{Sanchez-Lavega2014} with further points obtained in July 1992. Dotted lines indicate linear fits to the data. (b) All measurements of the Hexagon's vertices.Positions in CIRS images are highlighted with a red box, positions in Cassini ISS are highlighted with a yellow box, positions in ground-based images are highlighted with a green box. Dotted lines indicate polynomial fits to the data. The identification of vertices of the hexagon in 1980-81 and 1990-1991 was found after testing several permutations of the vertices and computing polynomial fits to the data. The case shown here is the one that simultaneously: (i) results in the best agreement between the instantanous drift rate of the hexagon from the polynomial fits here shown with the drift rates measured in different epochs, and (ii) has the minimal variations in the drift rate over time.  This last condition was verified by examining several permutations of the vertices identifications in the three sets of Voyager, HST and Cassini and later ground-based data.}   
\label{fig:hexagon_in_time}
\end{figure}

Figure \ref{fig:hexagon_in_time} shows a compilation of historical measurements of Saturn's hexagon. The figure shows details of the new measurements for 1990-1991 and a reconstruction of the positions of the vertices of the hexagon in time since 1980. The small positive drift rates in 2007-2011 combined with  negative drift rates in 1990-1991 and 1980-1981 in table \ref{tab:Hexagon_drifts} imply a complex variation of the hexagon's drift rate in time. We considered several possible identifications of the vertices of the hexagon and computed a 7$^{th}$ order polynomial fit to each vertex. This polynomial was used to compute the position of each vertex as a function of time and derive a mean drift rate as a function of time. This derived drift rate should match the drift rate measured in each individual period. Several possible identifications of the vertices of the hexagon are possible when joining measurements obtained in the 1980s, 1990s and 2010s. The simplest solution is the one that minimizes variations in the drift rate of the hexagon over time. This identification is the one used for the reconstruction of the vertices positions shown in this figure and was obtained after extensive testing of the different combinations of the data.

\begin{figure}[htp]
   \centering
   \includegraphics[width=12.50cm]{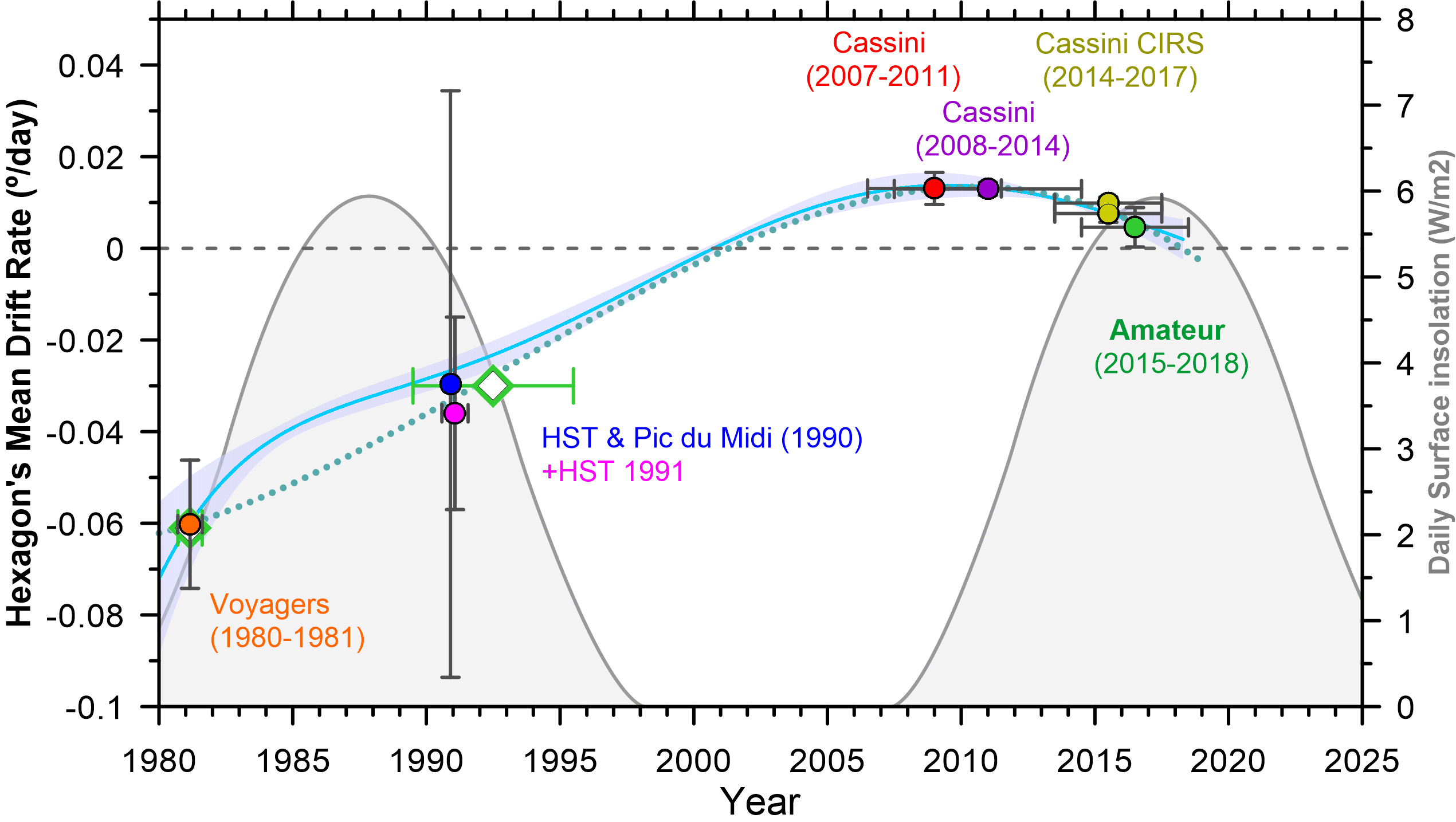}
   \caption{Hexagon drift rates in different years with measurement error bars (circles) and drift rates of the North Polar Spot (diamonds). A 3rd order polynomial fit to the hexagon drift rate is shown (dotted-line). We also show the drift rate as a function of time obtained from the polynomial fits to the hexagon's vertices positions assuming the identification of vertices in Figure \ref{fig:hexagon_in_time}. Error-bars here come from differences between the polynomials of the six vertices. The grey filled line shows solar irradiation at the hexagon's latitude as a function of time.}   
\label{fig:Hexagon_drift_rates}
    \end{figure}

Figure \ref{fig:Hexagon_drift_rates} shows the measured drift rates appearing in table \ref{tab:Hexagon_drifts} and the continuous drift rate calculated from the fit to the hexagon's vertices assuming the previous identification. This is compared with a polynomial fit to the drift rates measured in specific years  that naturally approaches very closely to our proposed solution. We also show the daily solar irradiation at the top of the atmosphere at the hexagon's latitude to display the seasonal radiative forcing at this latitude. The comparison of the different drift rates found in Voyager and Cassini data for the same season implies that the variation in the hexagon's drift rate is not seasonal. The figure also shows the drift rates of the NPS in the 1980s and 1990s. 


The hexagon is located at the eastward jet and is visible at the cloud level bellow the hazes in Cassini ISS images, in deeper levels in Cassini VIMS data and in Cassini CIRS thermal data of the stratosphere since 2014. Although seasonal effects at the upper levels are important, at the lower levels they should be minimal. For instance, the fast eastward jet where the hexagon is located has remained stable since the Voyager 1 in 1980 imaged this area (see \citealt{Antunano2018} and references therein). Certainly, seasonal changes have been observed in the upper troposphere and stratosphere, with temperatures in the tropopause changing from $\sim$78 to 91 K from 2004 to 2017 \citep{Fletcher2018NatCo}. However, the lower troposphere seems stable. Quantitative assesments of changes in the aerosol properties of the polar atmosphere in different periods of time are not available, although specific studies of the overall structure of the polar aerosols conclude that the base of the tropospheric haze in Saturn's north polar atmosphere is located at around 600 mbar and descends towards the polar region \citep{Sanz-Requena2018}. This implies clouds located below the levels affected by seasonal variations.

The survival of the hexagon during the polar night, the stability of the polar jet to the varying solar irradiation, the indirect determination of winds extending down at least to the $\sim10$ bar \citep{Sanchez-Lavega_Nature2011} level, and the recent determination of deep winds in Saturn's mid to low latitudes \citep{Galanti2019} convincingly agree that the polar jet extends deep in the atmosphere, well below levels affected by seasonal variations. Still, the hexagon's drift rate changes in time and the presence of the NPS and coincident drift rate in 1980-1981 and 1990-1991 periods, and its absence since the first images of the north polar area in 2007 seem to indicate the importance of this feature in modifying the hexagon's drift rate.

\section{Summary and conclusions}
With the demise of the Cassini mission in 2017, new observational data sets are needed to advance in our knowledge of Saturn's atmosphere. While HST remains as the main observational facility to study Saturn's atmosphere dynamics, we have showed the value of combining HST images with frequent ground-based observations provided by small telescopes operated by amateur astronomers. These observations can be used not only to follow the development of major storm systems \citep{Sanchez-Lavega2004, Sanchez-Lavega2019}, but also to survey the global atmospheric activity of the planet. We now summarize our main findings.

\begin{itemize}

\item The long-term tracking of a bright white spot at $6.7\pm1.6^{\circ}$N discussed by \citet{Sanchez-Lavega_NatCom2016} is extended with respect to data in that publication from 2014-2015 to 2014-2018. This long-lived white spot moved coherently over this period of time with velocities 
that accelerated from $444.3\pm3.1$ ms$^{-1}$ in 2014 to $452.4\pm1.7$ ms$^{-1}$ in 2018. These values are closer to the equatorial zonal winds at cloud level at the time of the Voyagers than during the Cassini mission.  Zonal winds obtained in Cassini VIMS images sensitive to the $\sim2$ bar level \citep{Choi2009,Studwell2018} are closer to the speed of the bright white spot than winds in the upper layers measured in Cassini ISS images. However, the white spot moves slightly faster than the deep zonal wind profiles. Winds obtained at the visible cloud at this latitude by \citet{Sanchez-Lavega_NatCom2016}  over HST images acquired in 2015 indicate that local equatorial winds outside the region of the white spot are similar to those obtained by Cassini. The interpretation in  \citet{Sanchez-Lavega_NatCom2016} that the bright white spot is probably a deeply rooted atmospheric system moving at zonal speeds characteristics of levels bellow the typically observed levels in visible images of the Equatorial Zone  still holds, but it suggests Cassini winds of the deep atmosphere measured with the VIMS instrument might require further detailed analysis in the future. 

\item A similar bright and fast feature has been found in the South Equatorial Zone at $6.7\pm2.2^{\circ}$S during most of 2018. Except for its elongated shape, this South equatorial bright spot is very similar in size, brightness, drift rate and closeness to the deep winds obtained by VIMS. 

\item  Cassini zonal winds in 2004-2009 and 2014 changed in the two layers sensed by the continuum band and methane band filters sensitive to the lower clouds and upper hazes respectively. Previous indications of equatorial zonal winds changing over the course of the Cassini mission \citep{Li2011}  and the equatorial thermal oscillations observed by the CIRS instrument  \citep{Fouchet2008, Orton2008, Guerlet2018} suggest that a thorough analysis of zonal winds with the best temporal resolution should be pursued with Cassini archived data complemented with cloud correlation analysis on current and future HST data.

\item Several equatorial clouds were tracked in September-October 2018 in the North and South latitudes of the EZ moving at intermediate speeds between those of the bright white spots and those of zonal winds at the cloud level. Features with similar intermediate velocities where also found in 2015 with HST data in the North EZ but not in the South\citep{Sanchez-Lavega_NatCom2016}. We interpret these varied values of velocities as a consequence of the vertical cloud structure and vertical wind shear that should be similar in the period 2015-2018.

\item A peculiar cloud feature in the northern edge of the EZ and observed from  2016  to 2018 moved 35 ms$^{-1}$ slower than the zonal winds. The morphology of this feature  in filters penetrating below the upper hazes  corresponds to a sheared structure. The zonal winds at the latitude of this feature ($15.2\pm1.0^{\circ}$) show a characteristic kink in their latitudinal structure with a local minimum in the relative vorticity and a local maximum in the meridional gradient of relative vorticity  that could cause a dynamical instability. The 35 ms$^{-1}$ slower velocity of this feature could be a manifestation of a phase speed of the wave disturbance.

\item Many long-lived features were regularly present at Saturn's tropical latitude at around 30$^{\circ}$N. All of them closely follow the zonal winds. Their longevity and characteristics imply a non convective nature and a comparison with Cassini ISS images and their position with respect to the zonal winds suggests they are a chain of cyclones moving at very close latitudes. 

\item The large AV that formed after Saturn's GWS in 2011 is still present in the planet but with a reduced contrast with the atmosphere. We have presented a summary of its properties since it formed in 2011. The vortex has moved significantly in latitude experiencing changes in its drift rate that are broadly consistent with Cassini zonal winds. Its drift rate is compatible with zonal winds slightly larger than those measured in 2009. This is in agreement with the zonal winds measured after the fade of the GWS \citep{Sayanagi2013}, and those predicted from thermal wind arguments \citep{Achterberg2014, Fletcher2016}.

\item Bright features at about 45$^{\circ}$N are present in many ground-based images of the planet. They are most probably related with the activity of the ribbon wave at 47$^{\circ}$N. The determination of their motions is compatible with either clouds moving with the zonal winds, or the phase speed of the ribbon wave.

\item Subpolar vortices like the isolated dark vortex at 60$^{\circ}$N and the ACA system at 64$^{\circ}$N were still active in the planet in 2018. For the ACA system this implies a coherent lifetime of the three vortices of at least 7 years,  only slightly smaller than the AV. The interactions they experienced in 2015 that resulted in a subpolar planetary scale perturbation \citep{delRio-Gaztelurrutia2018} has not developed again.

\item We extend the measurements of Saturn's hexagon drift rate in time showing changes from the measurements obtained by the Voyagers and HST in the 1980s and 1990s compared with a more stable and smaller drift rate during the Cassini mission. The long-term variations of this small drift rate are not in agreement with a simple seasonal variation and reinforce the role of the NPS in forcing the hexagon at the time of its discovery and the decade after. While the hexagon does not require a nearby polar anticyclone to maintain its stability, nearby large anticyclones like the NPS may interact with it modifying its drift rate. Such interactions might be modulated by the unknown depths of both the NPS and the hexagon.
\end{itemize}

\section*{Acknowledgments:} We thank two anonymous reviewers for constructive comments that enhanced the physical interpretation of the data presented. We are very thankful to the ensemble of observers operating small telescopes that submitted their Saturn observations to databases such as PVOL and ALPO Japan. Observations were also obtained from the following observatories: Pic du Midi in France and Centro Astron\'omico Hispano Alem\'an (CAHA) at Calar Alto, Spain. Saturn observations at the Pic du Midi observatory in 2017 were acquired by the Pic-Net team: F. Colas, M. Delcroix, E. Kraaikamp, R. Hueso, D. Peach, C. Sprianu, G. Therin, with funding from Europlanet 2020 RI, which has received funding from the European Union’s Horizon 2020 research and innovation programme under grant agreement No. 654208. At the time of the observations here reported the Centro Astron\'omico Hispano Alem\'an (CAHA) at Calar Alto was operated jointly by the Max Planck Institut f\"ur Astronomie and the Instituto de Astrof\'isica de Andaluc\'ia (CSIC) and is now operated by the Junta de Andaluc\'ia and the Instituto de Astrof\'isica de Andaluc\'ia. This work used data acquired from the NASA/ESA HST Space Telescope, associated with OPAL program (PI: Simon, GO13937), and archived by the Space Telescope Science Institute, which is operated by the Association of Universities for Research in Astronomy, Inc., under NASA contract NAS 5-26555. All HST Saturn maps are available at \url{http://dx.doi.org/10.17909/T9G593}. This work was supported by the Spanish MINECO project AYA2015-65041-P (MINECO/FEDER, UE), Grupos Gobierno Vasco IT-765-13, UPV/EHU UFI11/55 and ‘Infraestructura’ grants from Gobierno Vasco and UPV/EHU. A.A.S. and M.H.W. were supported by grants from the Space Telescope Science Institute, which is operated by the Association of Universities for Research in Astronomy, Inc., under NASA contract NAS 5-26555. A.A. and L.N.F. were supported by the European Research Council Consolidator Grant under the European Union's Horizon 2020 research and innovation program, grant agreement 723890. E.G.M. was supported by the Serra Hunter Programme, Generalitat de Catalunya.
\bibliography{mybibfile}

\end{document}